\documentclass[12pt,preprint]{aastex}

\shorttitle{Radio Cores of Bipolar PNs}
\shortauthors{Lee, Lim \& Kwok}

\begin{document}

\title{Optically Thick Radio Cores of Narrow-Waist Bipolar Nebulae}

\author{T.-H. Lee} 
\affil{National Optical Astronomy Observatory, Tucson, AZ 85719\\ and
  \\ Departemnt of Physics and Astronomy, University of Calgary,
  Calgary, AB T2N 1N4, Canada} \email{thlee@noao.edu}

\author{J. Lim}
\affil{Institute of Astronomy \& Astrophysics, Academia Sinica, Taipei
  106, Taiwan}
\email{jlim@asiaa.sinica.edu.tw}

\and

\author{S. Kwok} 

\affil{Department of Physics, Faculty of Science, University of Hong
Kong, Hong Kong \\ and \\ Departemnt of Physics and
Astronomy, University of Calgary, Calgary, AB T2N 1N4, Canada}
\email{sunkwok@hku.hk}

\begin{abstract}

  We report our search for optically thick radio cores in sixteen
  narrow-waist bipolar nebulae.  Optically thick cores are a
  characteristic signature of collimated ionized winds.  Eleven
  northern nebulae were observed with the Very Large Array (VLA) at
  1.3~cm and 0.7~cm, and five southern nebulae were observed with the
  Australia Telescope Compact Array (ATCA) at 6~cm and 3.6~cm.  Two
  northern objects, 19W32 and M~1-91, and three southern objects,
  He~2-25, He~2-84 and Mz~3, were found to exhibit a compact radio
  core with a rising spectrum consistent with an ionized jet.  Such
  jets have been seen in M~2-9 and may be responsible for shaping
  bipolar structure in planetary nebulae.

\end{abstract}


\section{Introduction}
\label{sec:intro-jet}

  Planetary nebulae (PNs) show a remarkably broad range of often
  intricate morphologies.  Butterfly nebulae, classified as bipolar
  nebulae with a narrow pinched waist at the center \citep{balick02},
  are probably the most spectacular, and yet at the same time most
  puzzling.  Under the classical interacting stellar winds (ISW) model
  \citep{kwok78}, a PN is formed when the fast ionized wind from its
  central star overtakes the material ejected earlier by the slow
  neutral wind and compresses the material into a dense shell.  This
  produces a round PN if both winds are isotropic.  A dense equatorial
  torus, which blocks the fast wind in that direction, is added to the
  interacting winds system to explain how asymmetric structure in PNs
  (e.g., bipolar) can be developed \citep{kah85, bal87}.  This,
  however, can only produce bipolar nebulae with wide waists, not
  those with narrow ``pinched'' waists.  It has been suggested that
  the creation of the latter nebulae requires a collimated fast wind
  \citep[e.g.,][]{soker00}.  \citeauthor{soker00}'s model invokes a
  binary star system to produce a collimated fast wind.  The system
  consists of a mass-losing AGB star and a main-sequence or white
  dwarf companion.  The AGB wind captured by the companion forms an
  accretion disk around the companion that loads and drives a
  collimated fast ionized wind.

  The existence of collimated outflows (COFs) has been suggested by
  numerous observations, usually from the ionized gas images.  For
  example, the highly collimated lobes of some PNs and PPNs
  (proto-planetary nebulae) imply that their morphology is shaped by a
  collimated outflow \citep[see review by][and references
  therein]{kwok03a}.  The first discovery of strongly collimated
  high-velocity bipolar outflows in a PN was by \cite{gieseking85}, in
  NGC~2392, the Eskimo nebula.  They analyzed from spectroscopic
  observations of this nebula, particularly in the [N~{\footnotesize
  II}] $\lambda$6583 emission line, they found a set of bipolar jets
  with a expansion velocity of $\sim$ 200~km s$^{-1}$ and a small
  angle of collimation less than 10$^{\circ}$.  Although unexpected at
  first, the pervasiveness of COFs in PNs is now widely recognized.
  COFs have diverse characteristics.  They are prominent in emission
  lines such as [N~{\footnotesize II}] and [S~{\footnotesize II}] and
  are observed as bipolar, multipolar, point-symmetric and jet-like
  outflows.  The expansion velocities of these COFs range from several
  tens to a thousand km~s$^{-1}$ \citep[see reviews of][]{lopez02,
  lopez03}.

  In addition to ionized gas, COFs have also been seen in neutral gas.
  Several authors have made detailed studies on CO emission lines of
  individual PNs and PPNs \citep[e.g.,][]{forveille98, huggins00}.
  The velocities for these CO outflows are found to be much smaller
  than those of the ionized COFs.  \cite{bujarrabal01} studied the CO
  emission from a large sample of PPNs as well as two yellow
  hypergiants and one young PN.  They reported collimated fast
  outflows in 28 of 32 objects, indicating that such outflows are a
  common phenomenon in the early PN stage.

  In the model proposed by \cite{soker00}, the collimated winds
  responsible for shaping butterfly nebulae are formed and collimated
  close to the central star system.  They are difficult to observe in
  optical wavelengths directly, because dust around the equator limits
  the depth into the center that optical observations can probe.  The
  dust is optically thin to radio emission, which can therefore probe
  deep into the center to look for central jets, if the jets are at
  least partially ionized.  Using the VLA, \cite{lim00,lim03} have
  indeed found a collimated wind in the form of a pair of ionized jet
  in the core region of the narrow-waist bipolar nebula M~2-9.

  M~2-9 has a bright central core and two cigar-shaped lobes extending
  from opposite sides of the core $\sim 20''$ to the north and south.
  The lobes of the nebula contain three pairs of symmetric knots or
  bright features: two of them are with mirror symmetry and one pair
  with point/mirror symmetry.  Because of its spectacular morphology,
  M~2-9 is one of the most well-studied planetary nebulae.  Since its
  discovery in 1947, the core and the outline of the lobes have
  remained stationary, but the knots and bright features have moved
  laterally from the west to the east \citep[e.g.,][]{allen72,
  kohoutek80}.  Various models have been proposed to explain the
  lateral motions of bright features of M~2-9.  For example,
  \cite{allen72} proposed that the changes were the result of a
  rotating beam of ionizing radiation from the star shining through a
  matched pair of holes in orbiting clouds.  \cite{kohoutek80} noted
  that it is unlikely such model can result in positional changes only
  in the east-west direction, and instead suggested that the nebula
  was rotating as a whole.  However, since \cite{goodrich91} concluded
  that the distance of M~2-9 cannot be only 50 pc as proposed by
  \cite{kohoutek80}, the unreasonable large velocities of the knots
  have made their model problematic.
 
  Recently in explaining the structural changes of M~2-9,
  \cite{doyle00} and \cite{livio01} both invoke a symbiotic star
  system to produce a mirror-symmetric jet.  Direct evidence of such a
  jet is provided by observations of M~2-9 at 6~cm, 3.6~cm, 2~cm, and
  1.3~cm by \citep{lim03}.  Their observations were largely
  insensitive to the large-scale free-free emission from the nebula
  and focused instead on the compact radio core at the center found by
  \cite{kwok85}.  Their results indicate that this optically thick
  core has a spectral index of $0.67 \pm 0.01$, which is close to the
  spectral index of 0.6 expected from an isothermal outflow expanding
  at a constant velocity and opening angle \citep{reynolds86}.  More
  importantly, at the shortest wavelength of 1.3~cm where the jet is
  best resolved, its bipolar lobes show a mirror-symmetric structure
  \citep[Figure 1 of][]{lim03}.  Both lobes point in the directions of
  mirror-symmetric brightenings in the nebular walls seen in the {\it
  HST} image observed just 2 years earlier \citep[N1 and S1 in Figure
  2 of][]{doyle00}.  The qualitative properties of this jet are in
  good agreement with those predicted by \cite{livio01}.

  Motivated by the discovery of the ionized jets in M~2-9, we launched
  a search for collimated ionized winds in butterfly nebulae using the
  VLA and the ATCA.  We selected a total of 16 objects from Table 1 of
  \citet{soker00} to search for optically thick radio cores -- a
  characteristic signature of ionized jets.  The selected objects were
  classified by \cite{soker00} to have either narrow waists (10
  objects) or possibly narrow waists (6 objects), whose morphology
  cannot be determined with absolute certainty based on the optical
  images obtained by \cite{corradi95}.  We are interested in finding
  any central source that is optically thick at short wavelengths
  where the nebular material is already optically thin.  Our goal is
  to identify suitable candidates for follow-up observations with
  higher sensitivity and resolution to reveal the structure of the
  source.  We observe each nebula with two wavelengths to calculate
  its spectral index, which tells us if the emission is optically
  thick.  In this paper, we report the results of our search for the
  radio cores.  The observations and data reduction are described in
  Section \ref{sec:data-jet}.  In Section \ref{sec:results-jet}, we
  report that radio emission were detected in 9 sources, and that 5 of
  those detected have optically thick radio cores.  Section
  \ref{sec:discussion} discusses the implication of the high mass loss
  rates derived from the flux densities of the radio cores.  Lastly, a
  summary is given in Section \ref{sec:sum-jet}.

\section{Observations and Data Reduction}
\label{sec:data-jet}

  The survey sample comprises 11 northern objects (Declination $\geq
  -40^\circ$ for the VLA observation limits), IRAS~07131-0147, M~1-16,
  NGC~2818, NGC~6302, 19W32, HB~5, NGC~6537, M~3-28, M~1-91, M~2-48,
  and NGC~7026, observed with the VLA at 1.3~cm and 0.7~cm, and 5
  southern objects, He~2-25, He~2-36, He~2-84, Th~2-B, and Mz~3,
  observed with the ATCA at 6~cm and 3.6~cm.  The observing
  wavelengths are the shortest wavelengths feasible at the time of
  observations, which give the highest angular resolutions possible.
  The optical properties of the objects are listed in Table
  \ref{tb:objects-jet}, with columns 1 and 2 giving the common name
  and the PN G name based on galactic coordinates, columns 3 and 4
  giving Right Ascension and Declination, column 5 indicating if they
  appear to have a very narrow waist, column 6 stating if the PN has a
  compact central source visible in optical images, and columns 7
  indicating if the nebula shows point or mirror symmetric structure.
  These properties are mainly from \cite{soker00}, with some update
  obtained from recent optical images.  A question mark means the
  attribute cannot be determined from available images.

\subsection{VLA Observations}

  The 11 northern objects were observed at 1.3~cm and 0.7~cm with the
  A-configuration in 2002 March 10.  The A-configuration is the most
  extended configuration, hence it provide the highest angular
  resolution achieved with the VLA.  It allows us to not just look for
  but resolve if possible a central compact and optically think
  source, as well as to better discriminate against if not resolve out
  extended nebular emission.  The observations were conducted in
  fast-switching mode with a calibration cycle time of just 2-3
  minutes in order to correct for rapid phase variation at such short
  wavelengths.  The integration time for the individual sources is
  about 20 minutes each at 1.3~cm and 0.7~cm.  3C286 was observed for
  about 5 minutes in order to calibrate the flux scales for individual
  sources.  The data were reduced and analyzed using the package AIPS
  following the standard procedures for reducing high-frequency data
  outlined in Appendix D of the AIPS Cookbook \citep{nrao03}.  The
  calibrated data were then Fourier transformed, weighted and CLEANed
  to generate the final CLEAN images.  Natural weighting was applied
  in order to obtain images with minimum-noise levels.  We achieved an
  angular resolution of $\sim 0.08''$ at 1.3~cm and $\sim 0.04''$ at
  0.7~cm for the final CLEAN maps.

\subsection{ATCA Observations}

  The 5 southern objects were observed at 6~cm and 3.6~cm with the
  6-km ``6C'' configuration in 2003 March 15 and 16.  These two
  wavelengths were the shortest wavelengths feasible when the
  observations took place.  With 6-km being the most extended
  configuration, they provide the highest angular resolution
  achievable at the time of observation.  We observed each source for
  15 minutes roughly every 40 minutes over a period of 12 hours in
  order to optimize the {\it uv}-coverage, obtaining a total
  integration time of 4-5 hours on each object.  The primary
  calibrator 1934-638 was observed for flux calibration.  The data
  were reduced using the reduction package MIRIAD following the
  procedures outlined in Miriad Users Guide \citep{sault03}.  The
  final CLEAN images were obtained by performing inverse Fourier
  transforms from the calibrated visibility datasets to images, and
  then deconvolve the sidelobes using the CLEAN algorithm.  Natural
  weighting was used and we attained an angular resolution of $\sim
  3''$ at 6~cm and $\sim 1.5''$ at 3.6~cm.  Where applicable, the
  self-calibration was performed to improve the image quality.  In the
  end, only He~2-84 has strong enough emission to allow phase
  self-calibration.

\subsection{Optical Observations}

  Three objects, He~2-36, He~2-84, and 19W32, which have radio
  emission detected in our study but have no high resolution optical
  images, were observed with the 2.3m Telescope in Siding Spring
  Observatory (SSO) in 2003 May 26-30.  The Imager instrument and
  three narrow-band filters, H$\alpha$, [N~{\footnotesize II}] and
  [O~{\footnotesize III}] were used to take photometric images.  The
  center wavelength and bandwidth of each filter are listed in
  Table~\ref{tb:filter}, along with the best achieved FWHM during the
  observations.  These filters' bandwidths of $\le$ 10~\AA~allow
  minimize contaminations from adjacent emission lines.  The CCD has
  dimensions of 1024x1024, with a focal plane scale of 0.59$''$/pixel
  determined by the telescope optics.

  The images were reduced with standard techniques using IRAF software
  \citep{massey97}.  The averaged bias frames and the overscan level
  were subtracted from each target image, which was then trimmed to
  discard the overscan region and the first and last few columns and
  rows.  The dark current was negligible during the observations, so
  no dark current subtraction was performed.  Finally, the images were
  divided by the flat-field image for each filter.  The exposure
  frames of each object were then combined to generate the final
  images in each filter, using a statistical method that removes
  cosmic ray contamination.

\section{Results and Analyses}
\label{sec:results-jet}

  Among the 16 objects observed, we detected radio emission in nine of
  them, and five of which have a central compact source.  In Table
  \ref{tb:result} we list the detect limits (3$\sigma$) of our
  observations, and the type of emission that were detected.  A more
  detailed description of the detections and the analysis for the
  compact sources are given in the following subsections.

\subsection{Northern Survey}
\label{sec:norsurvey}

  Of the 11 objects observed by VLA, we found that NGC~6302, Hubble~5
  and NGC~6537 only exhibit extended radio emission at 1.3~cm that
  could not be properly mapped due to the lack of short baselines in
  our observations.  Two objects, 19W32 and M~1-91, show only a
  compact central source at both 1.3~cm and 0.7~cm.  No emission was
  detected for the remaining six objects.  This is the first time the
  centers of all these nebulae have been examined at such high angular
  resolutions at radio wavelengths.  Note that for the six objects with
  non-detection in our observations, some have been detected at 6~cm
  by \cite{zijlstra89} and \cite{aaquist90}, whose surveys are more
  sensitive to the extended emission.

\begin{figure}
\plotone{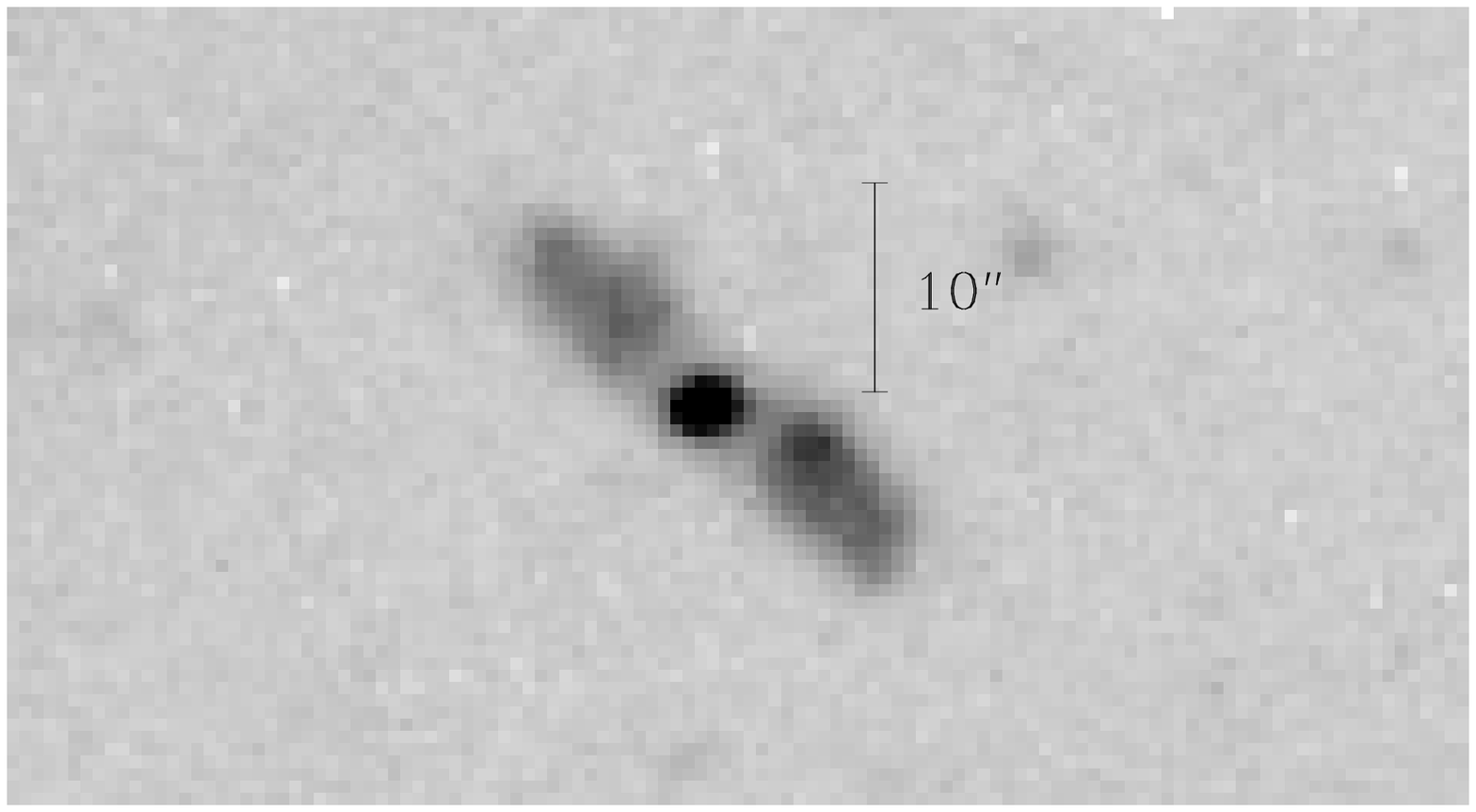}
\plottwo{f1b.eps}{f1c.eps}
\caption{Optical and radio images of 19W32.  a) SSO H$\alpha$ image in
        a linear scale.  b) VLA 1.3~cm (left) and 0.7~cm (right) radio
        continuum images with contours superposed.  The contours start
        at $3\sigma$, with the noise level $\sigma = 1.1 \times
        10^{-4}$ Jy/beam at 1.3~cm and $\sigma = 3.6 \times 10^{-4}$
        Jy/beam at 0.7~cm.  The radio images are on a much smaller
        scale, with the axis range only $\sim 1''$.  The position of
        the radio core coincides with the optical stellar core.  In
        this and subsequent figures, the ellipse in the box in the
        lower-left corner represents the half-power extent of the
        synthesized beam.}
\label{fg:19w32}
\end{figure}

\begin{figure}
\plotone{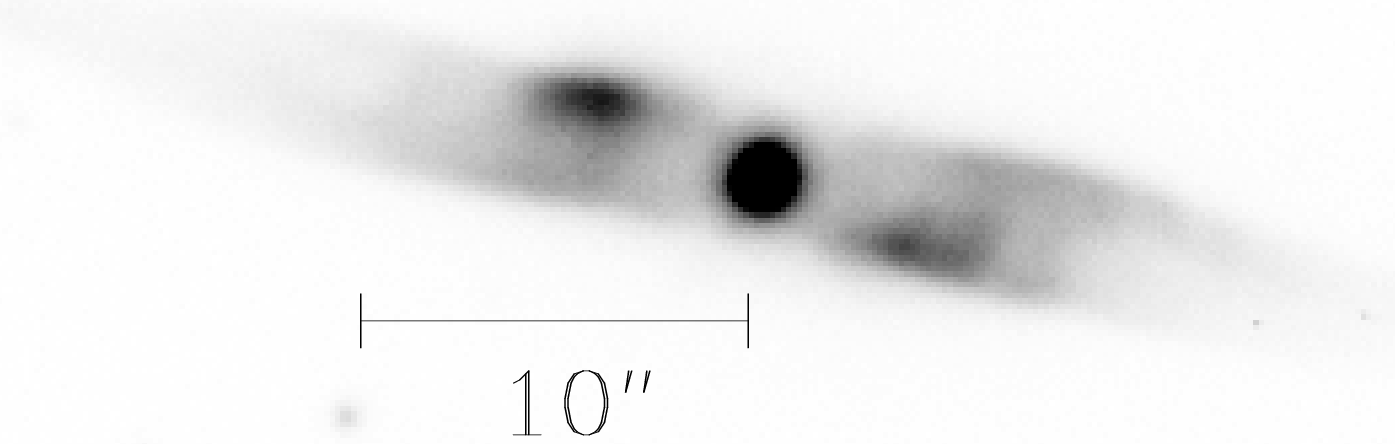}
\plottwo{f2b.eps}{f2c.eps}
\caption{Optical and radio images of M~1-91.  a) IAC H$\alpha$ image
        in a linear scale.  b) VLA 1.3~cm (left) and 0.7~cm (right)
        radio continuum image with contours superposed.  The contours
        start at $3\sigma$, with the noise level $\sigma = 9.95 \times
        10^{-5}$ Jy/beam at 1.3~cm and $\sigma = 2.7 \times 10^{-4}$
        Jy/beam at 0.7~cm.  The radio images are on a much smaller
        scale, with the axis range only $\sim 1''$.  The position of
        the radio core coincides with the optical stellar core.}
\label{fg:m1-91}
\end{figure}

  Here we present the images for the radio cores of 19W32 and M~1-91,
  along with their optical images (Figures~\ref{fg:19w32} and
  \ref{fg:m1-91}).  Both objects exhibit a compact stellar-like
  optical core.  We obtained the optical images of 19W32 from SSO 2.3m
  telescope observations as described above, those of M~1-91 from the
  IAC (Instituto de Astrof\'{i}sica de Canarias) catalog
  \citep{manchado96}.  All images are oriented with north up and east
  to the left.

\subsubsection{19W32}

  The H$\alpha$ and [N~{\footnotesize II}] images of 19W32 show a
  prominent stellar-like optical core and two lobes in opposite
  directions, extending $\sim 10''$ to the northeast and southwest.
  Only the H$\alpha$ image is presented in Figure~\ref{fg:19w32}.  No
  [O~{\footnotesize III}] emission was detected for this nebula.  The
  radio continuum images only show a compact core, which coincides
  with the optical stellar core.

\subsubsection{M~1-91}

  The H$\alpha$, [N~{\footnotesize II}] and [O~{\footnotesize III}]
  images of M~1-91 shows an optical core and two lobes in opposite
  directions, extending at least $20''$ to the northeast and
  southwest.  Since the images of the three filters show similar
  structures, only the H$\alpha$ image is presented in
  Figure~\ref{fg:m1-91}.  M~1-91's morphology is very similar to
  M~2-9, except that the brightenings of the lobes being point
  symmetric instead of mirror symmetric.  The radio continuum images
  only show a compact core, which coincides with the optical stellar
  core.  The 1.3~cm emission of M~1-91 shows some extended structures
  that might align with the point-symmetric structures.  This needs
  further observations to confirm.

\subsubsection{Spectral Indices of the Radio Cores}
\label{sec:tb}

  The spectral index, defined as the slope of the radio spectrum
  $\frac{d\log F_{\nu}}{d\log{\nu}}$, indicates the emission
  mechanisms and optical depths of the radio emission.  Optically
  thick free-free emission has a rising spectrum, optically thin
  free-free emission has a flat spectrum ($\sim -0.1$), and optically
  thin nonthermal emission has a steeply falling spectrum.  In the
  case of a spherically symmetric nebula, the radio emission has a
  spectral index of 0.6 ($S_{\nu} \propto \nu^{0.6}$) for an
  isothermal stellar wind with a constant expansion velocity
  \citep{wright75, panagia75}.  For a special case of collimated
  ionized winds with constant opening angle, the radio emission also
  has a spectral index of 0.6 \citep{reynolds86}.

  In order to determine the properties of the radio cores of 19W32 and
  M~1-91, we calculated the spectral indices of their radio emission.
  To determine the flux densities of the radio cores observed in 19W32
  and M~1-91, a two-dimensional Gaussian structure was fit to the
  measurements by applying the AIPS tasks IMFIT to the clean map and
  OMFIT to the visibility data.  Because of the low S/N ratio in these
  observations, neither task gives an obviously superior fit.
  Therefore results from both tasks are given.  The entire dataset
  were used for fitting at both 1.3 and 0.7~cm.  We also fit a more
  restricted dataset at 0.7~cm chosen to have the same sampling range
  in angular scales as at 1.3~cm to check for a greater inclusion of
  any extended nebula emission in the larger synthesized beam.  We
  found that both procedures gave the same results within measurement
  uncertainties ($< 1\sigma$).  Thus, in Table~\ref{tb:vla-flux}, only
  the results obtained by fitting the datasets over their entire {\it
  uv} range are listed.  The rising spectra of both 19W32 and M~1-91
  suggest that their radio cores are produced by optically thick
  free-free emission.  Fits to the measured visibilities show that
  these cores are resolved along one dimension; i.e., the minor axis
  of the synthesized beam.  The size of the emitting source will be
  discussed further in \S\ref{sec:mdot}
 
\subsection{Southern Survey}
\label{sec:sousurvey}

\begin{figure}
\plottwo{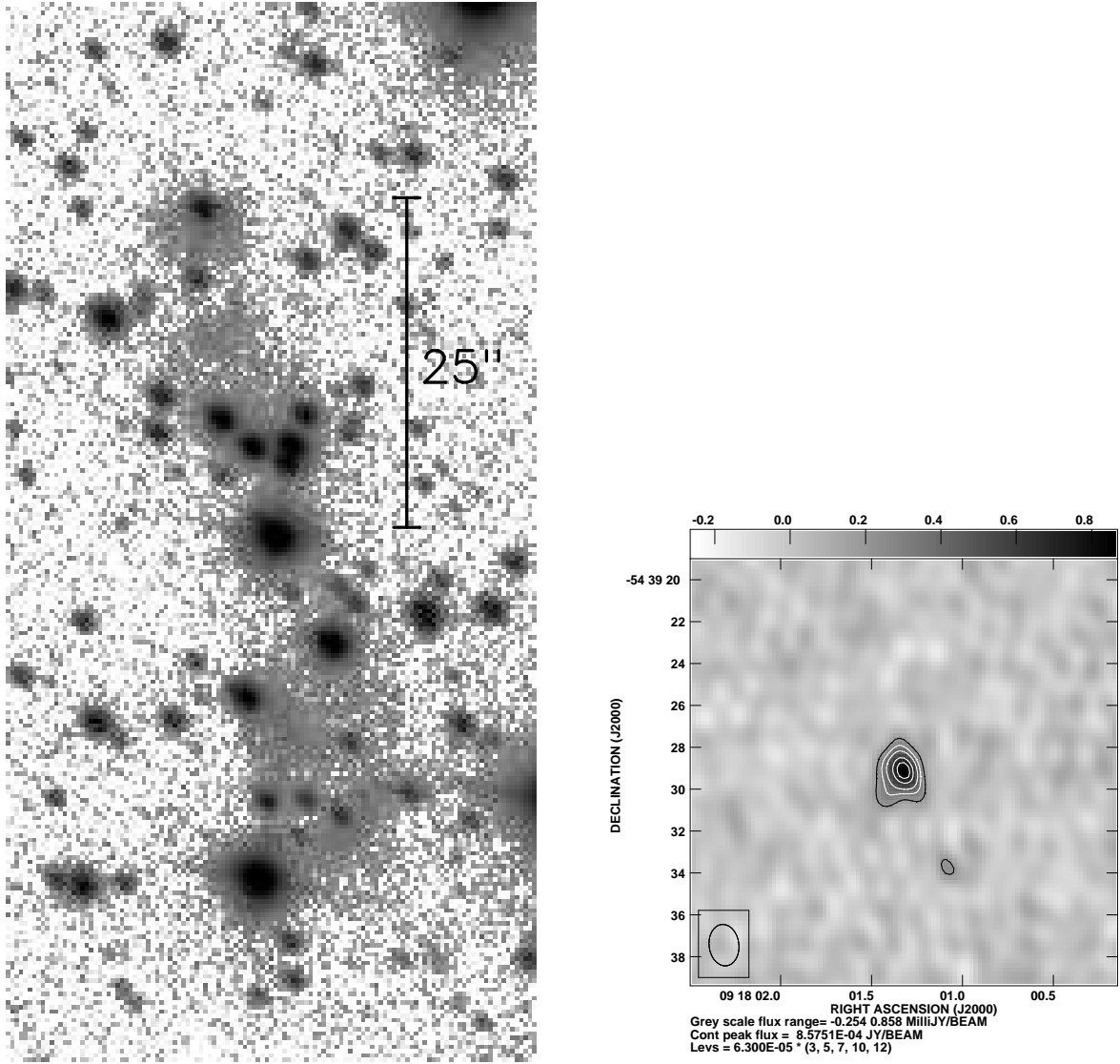}{f3b.eps}
\caption{Optical and radio images of He~2-25.  a) NTT R-band image
        \citep{lee07} in a logarithmic intensity scale.  b) ATCA
        3.6~cm (right) radio continuum image in greyscale with
        contours superposed.  The contours start at $3\sigma$ with the
        noise level $\sigma = 6.3 \times 10^{-5}$ Jy/beam at 3.6~cm.
        No emission is detected at 6~cm.}
\label{fg:he2-25}
\end{figure}

\begin{figure}
\plottwo{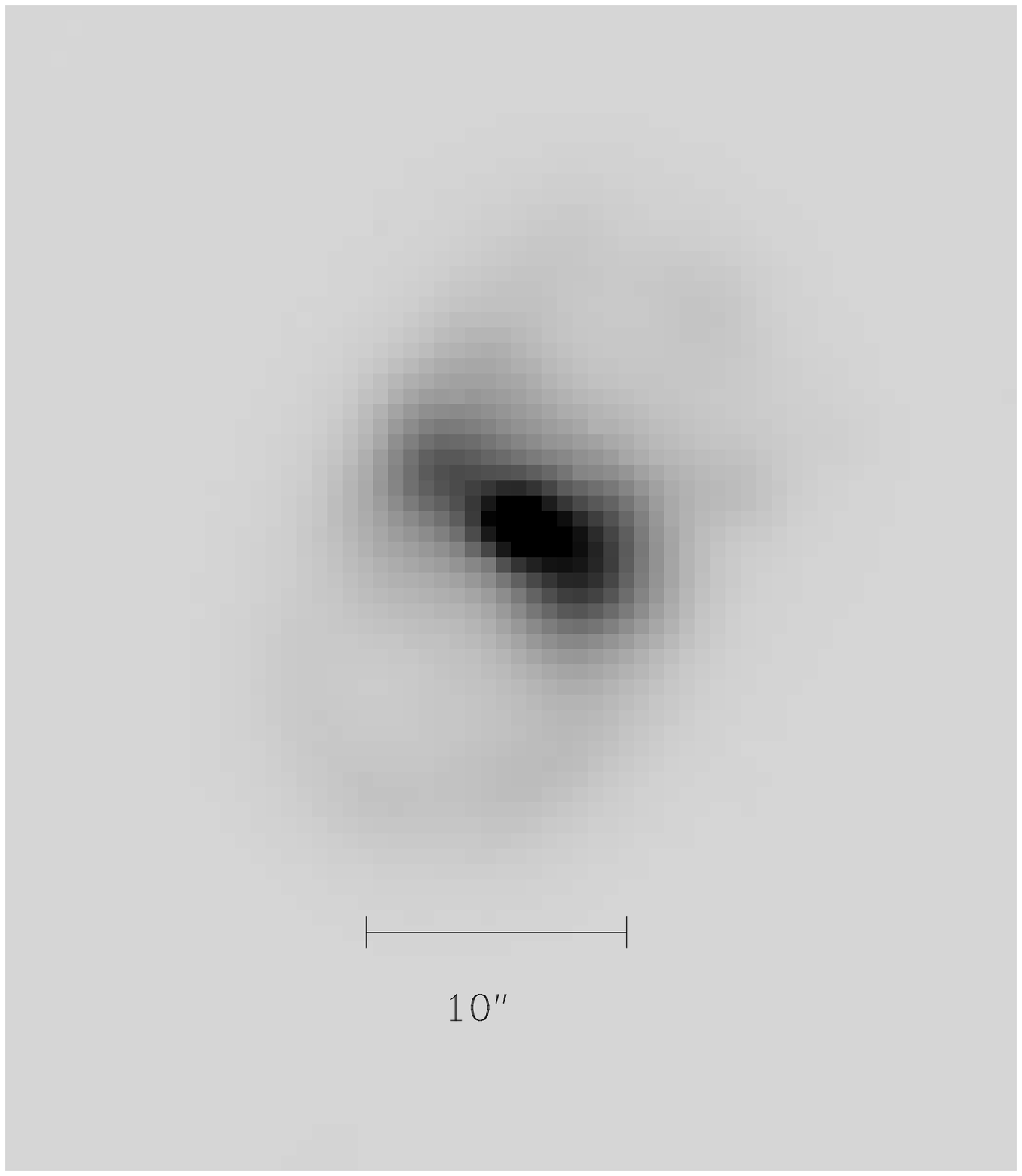}{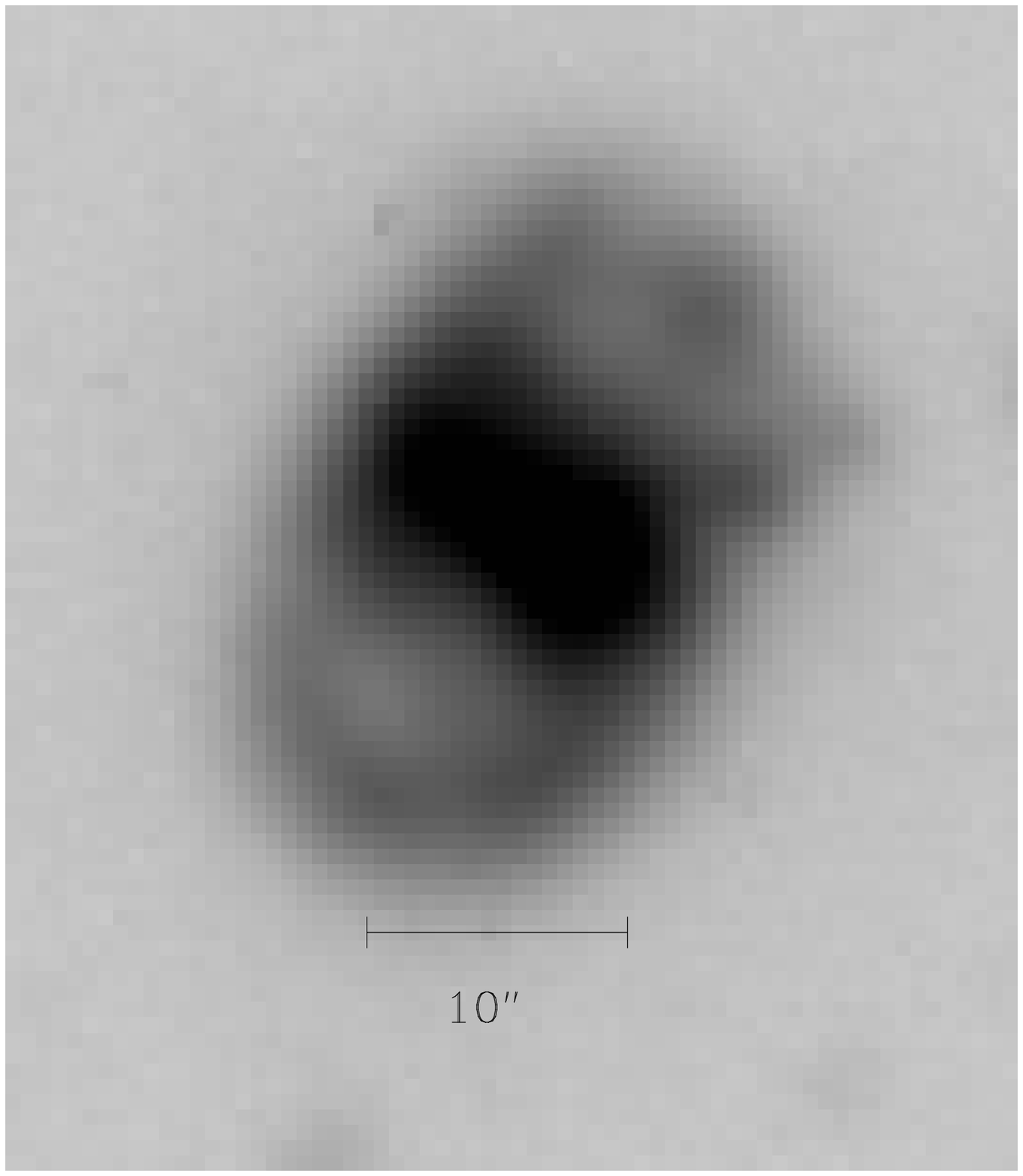}
\plottwo{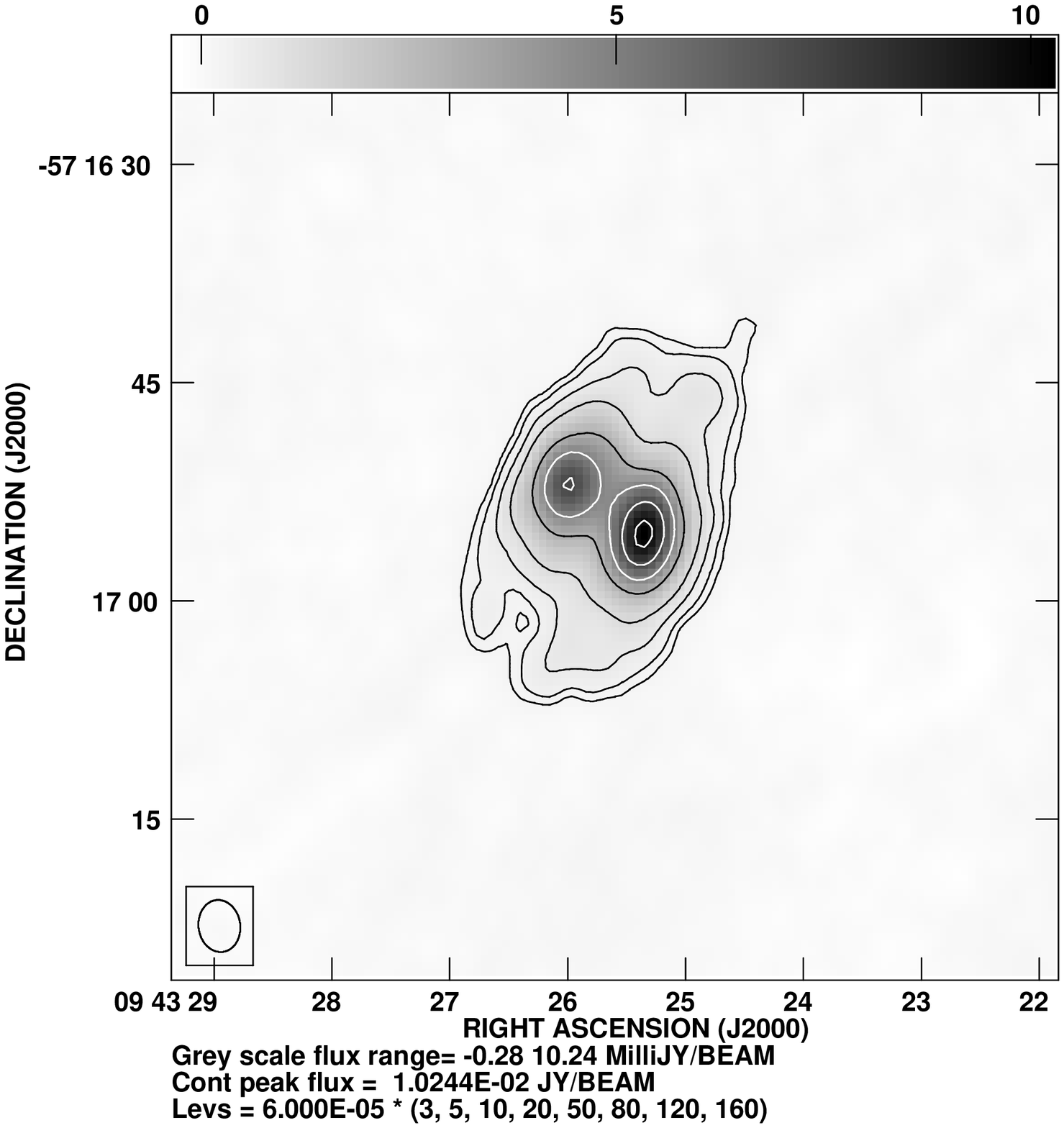}{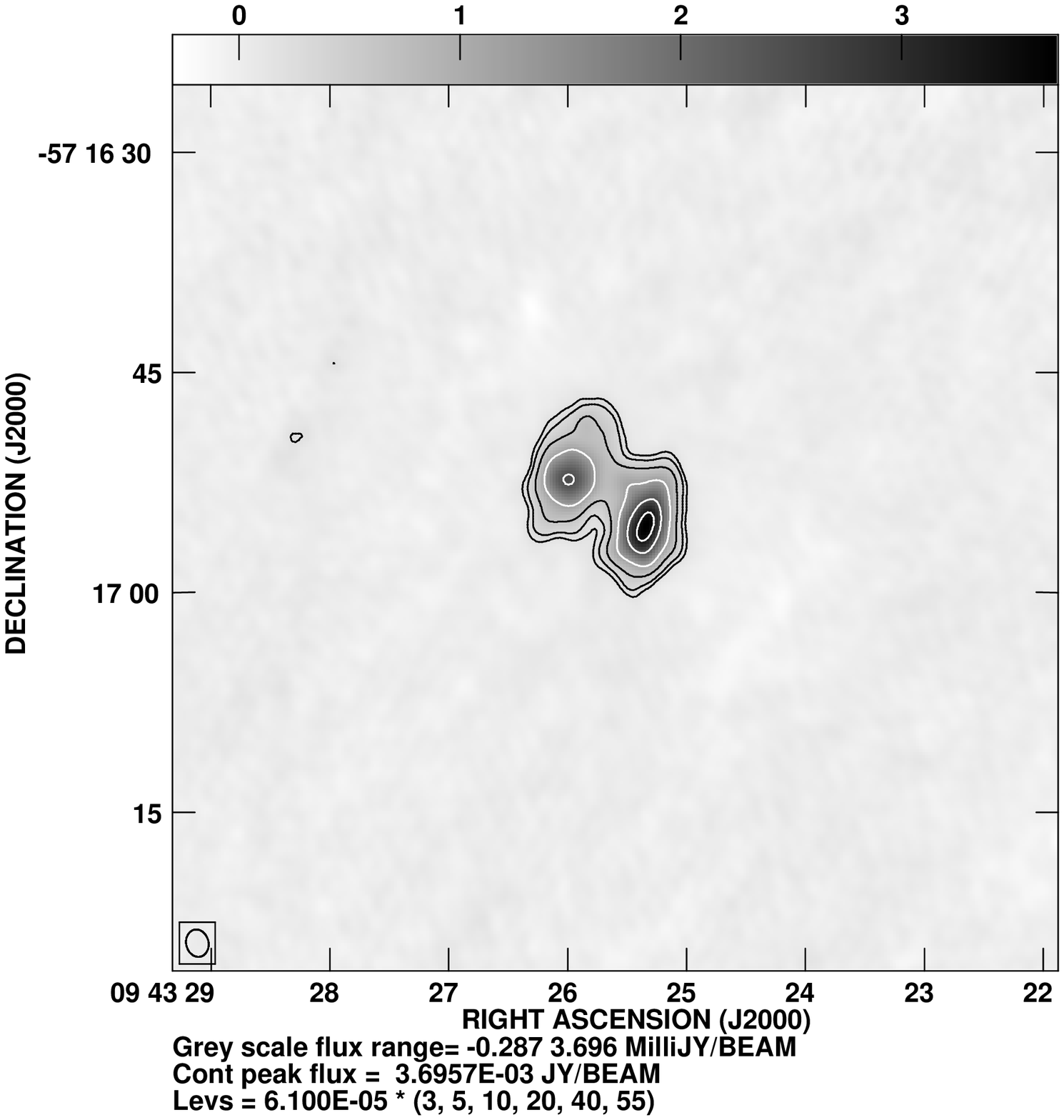}
\caption{Optical and radio images of He~2-36.  a) SSO H$\alpha$ images
        show the detailed structures of the center in a linear scale
        (left) and the outer lobes in a logarithmic intensity scale
        (right).  b) ATCA 6~cm (left) and 3.6~cm (right) radio
        continuum image in greyscale with contours superposed.  The
        contours start at $3\sigma$ with the noise level $\sigma = 6.0
        \times 10^{-5}$ Jy/beam at 6~cm and $\sigma = 6.1 \times
        10^{-5}$ Jy/beam at 3.6~cm.}
\label{fg:he2-36}
\end{figure}

\begin{figure}
\plottwo{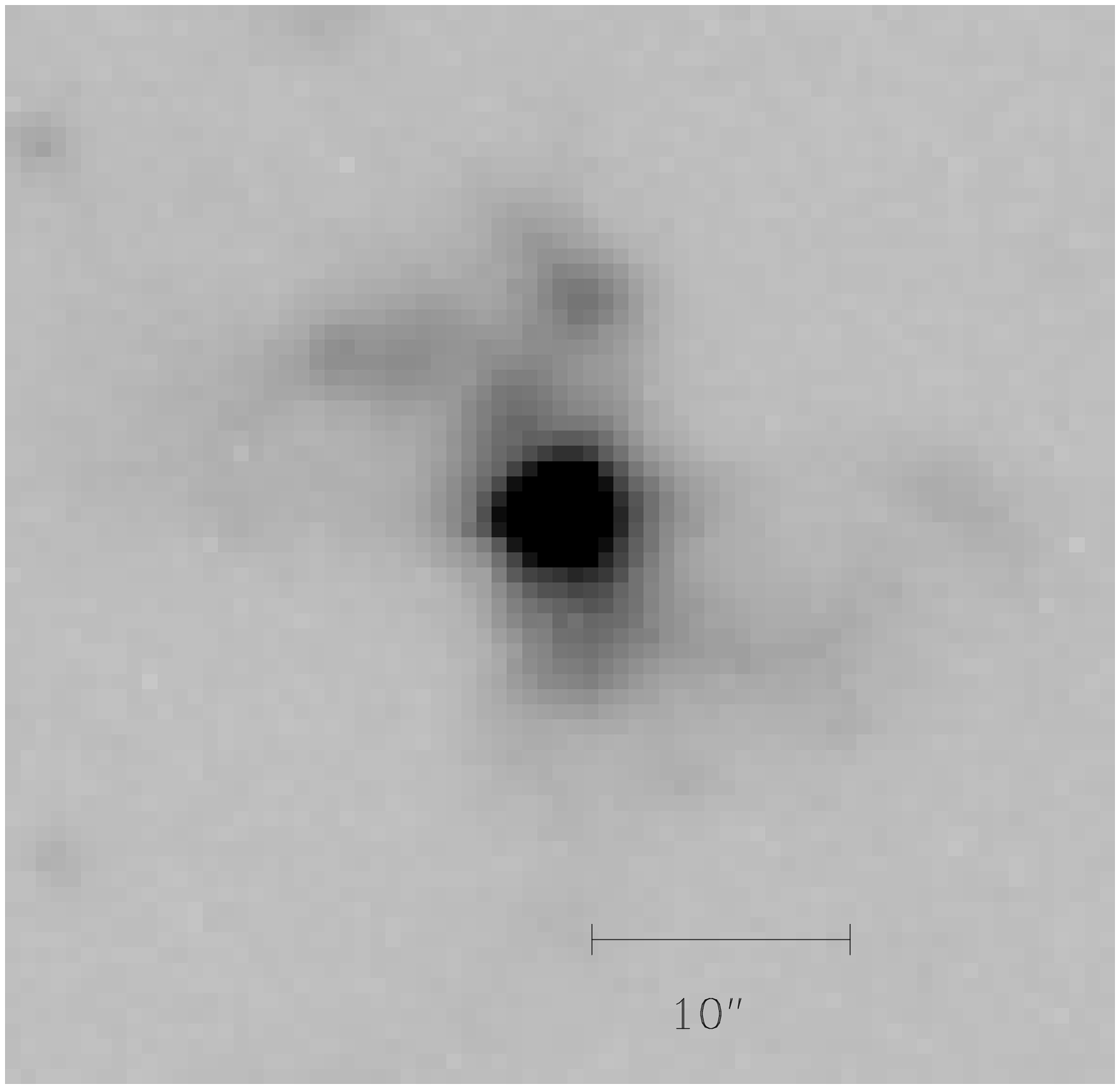}{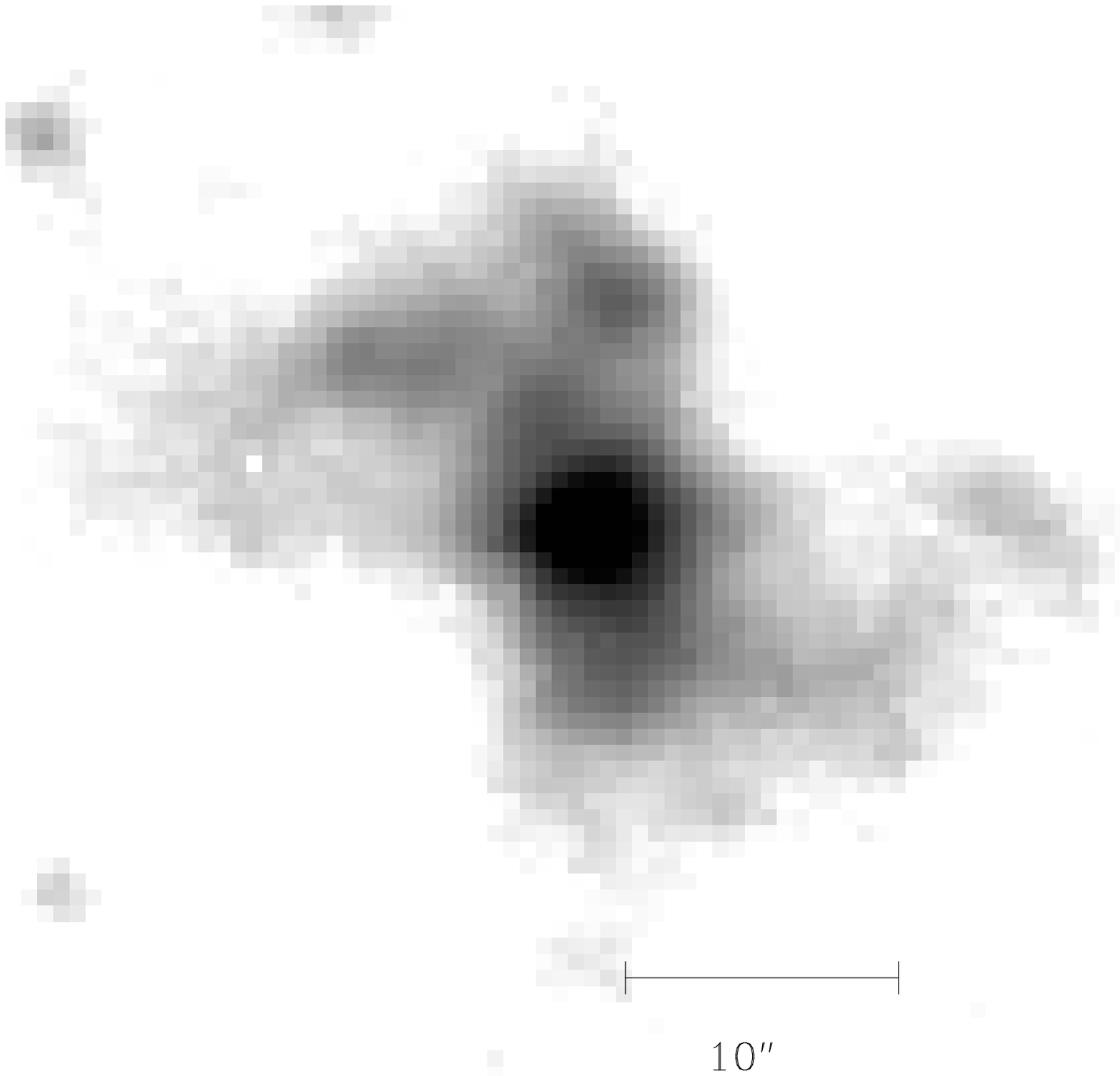}
\plottwo{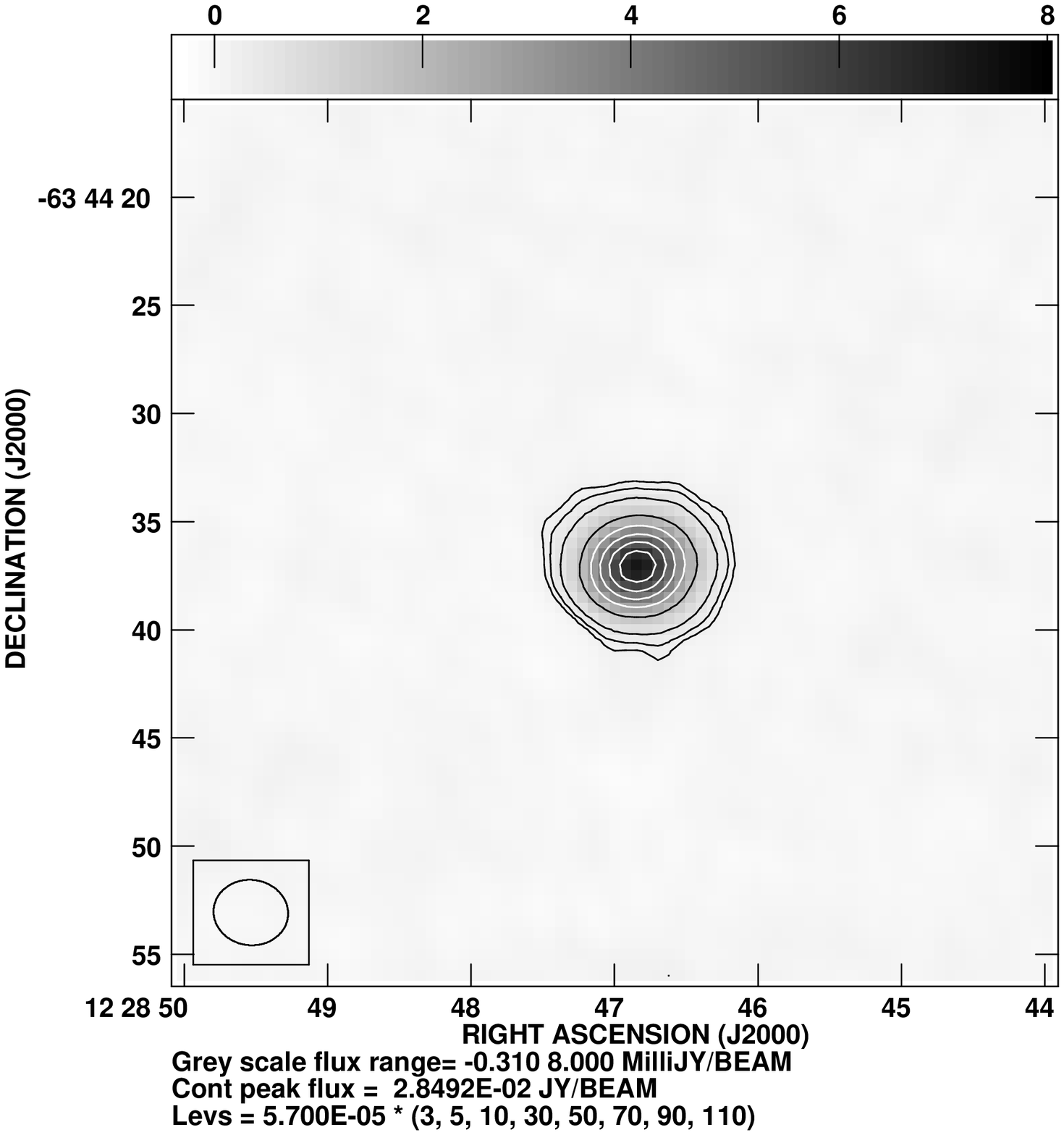}{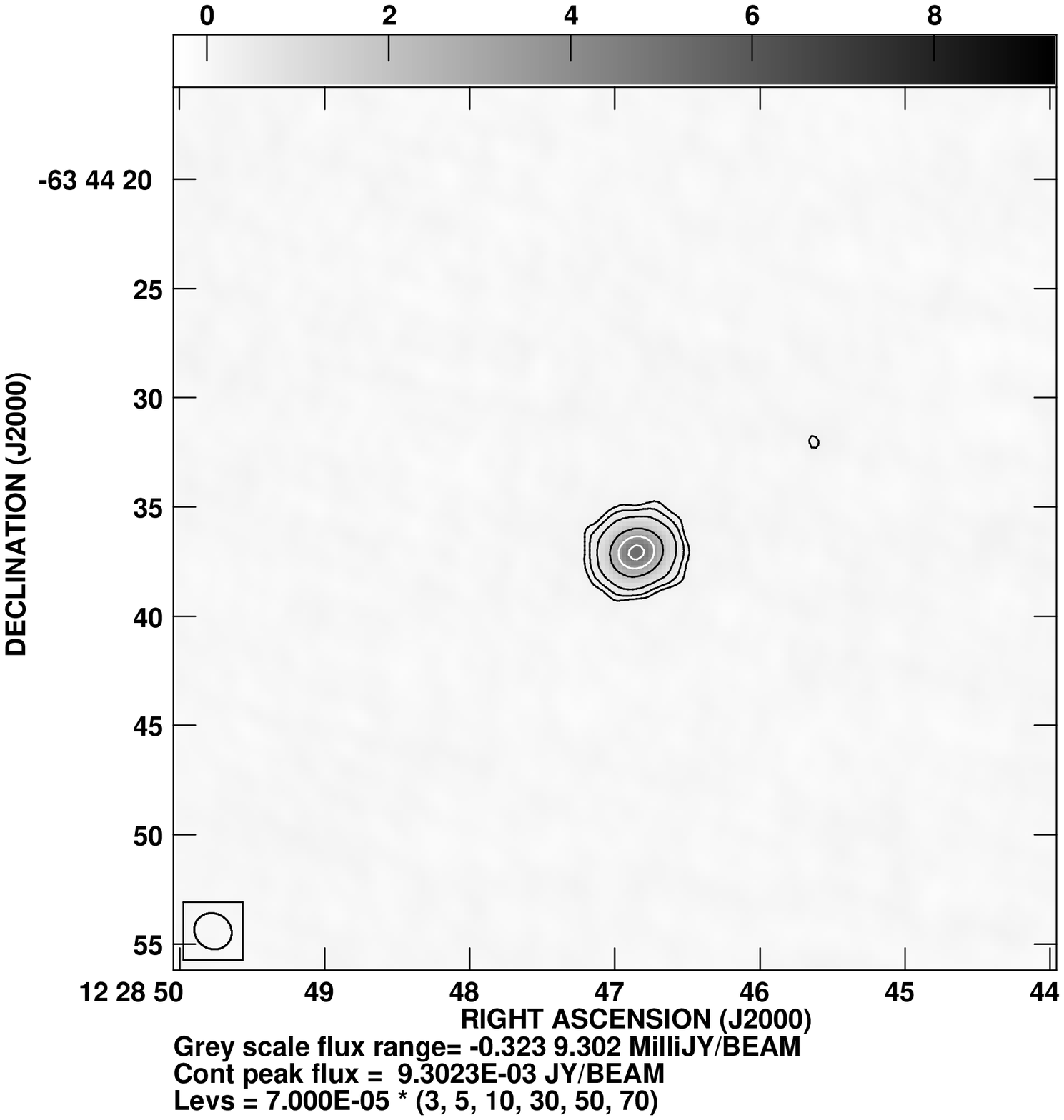}
\caption{Optical and radio images of He~2-84.  a) SSO H$\alpha$ images
        show the detailed structures of the center in a linear scale
        (left) and the outer lobes in a logarithmic intensity scale
        (right).  b) ATCA 6~cm (left) and 3.6~cm (right) radio
        continuum image in greyscale with contours superposed.  The
        contours start at $3\sigma$ with the noise level $\sigma = 5.7
        \times 10^{-5}$ Jy/beam at 6~cm and $\sigma = 7.0 \times
        10^{-5}$ Jy/beam at 3.6~cm.  The position of the radio source
        coincides with the optical stellar core.}
\label{fg:he2-84}
\end{figure}

\begin{figure}
\begin{center}
\includegraphics[width=40mm]{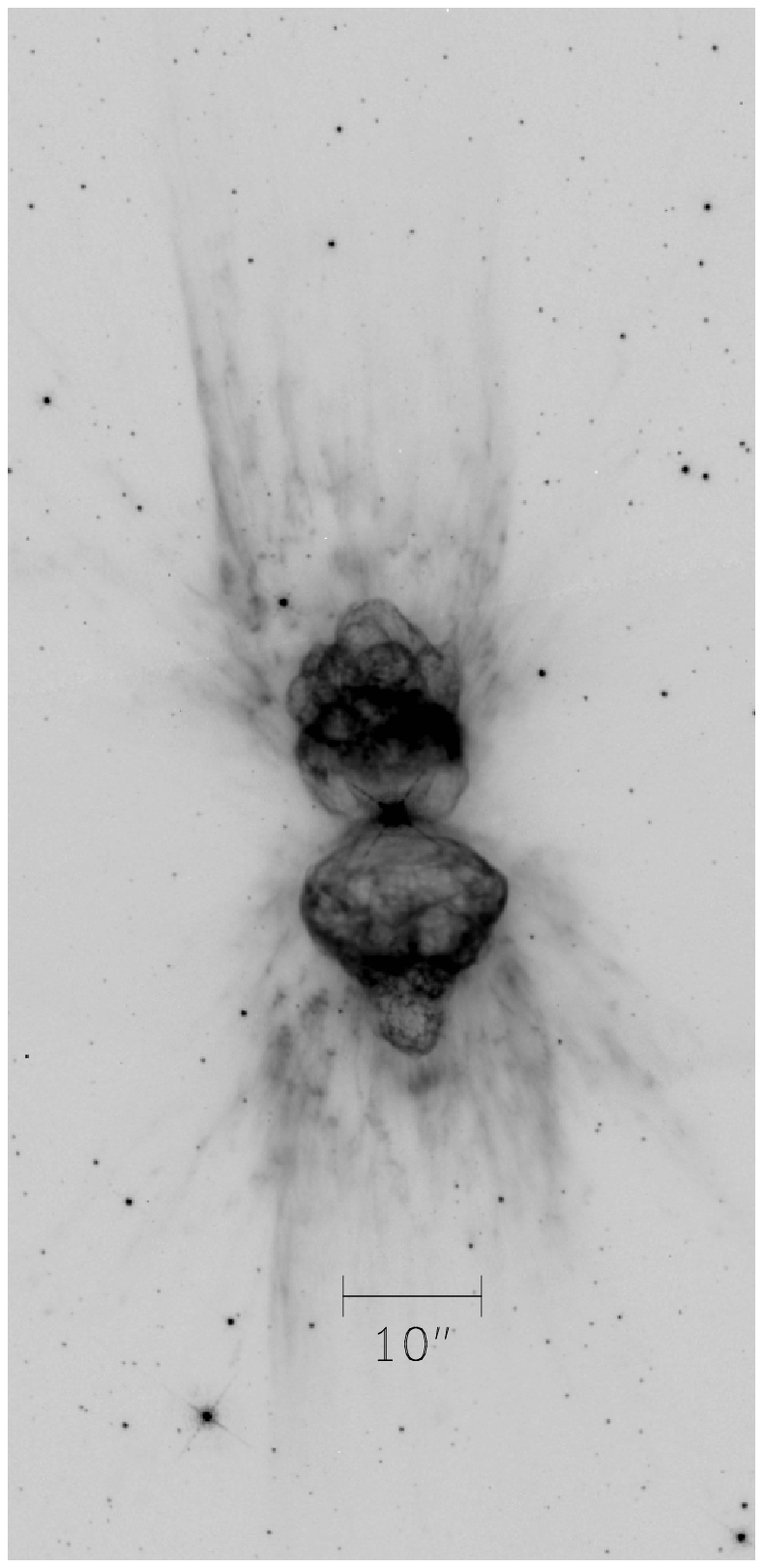} \\
\plottwo{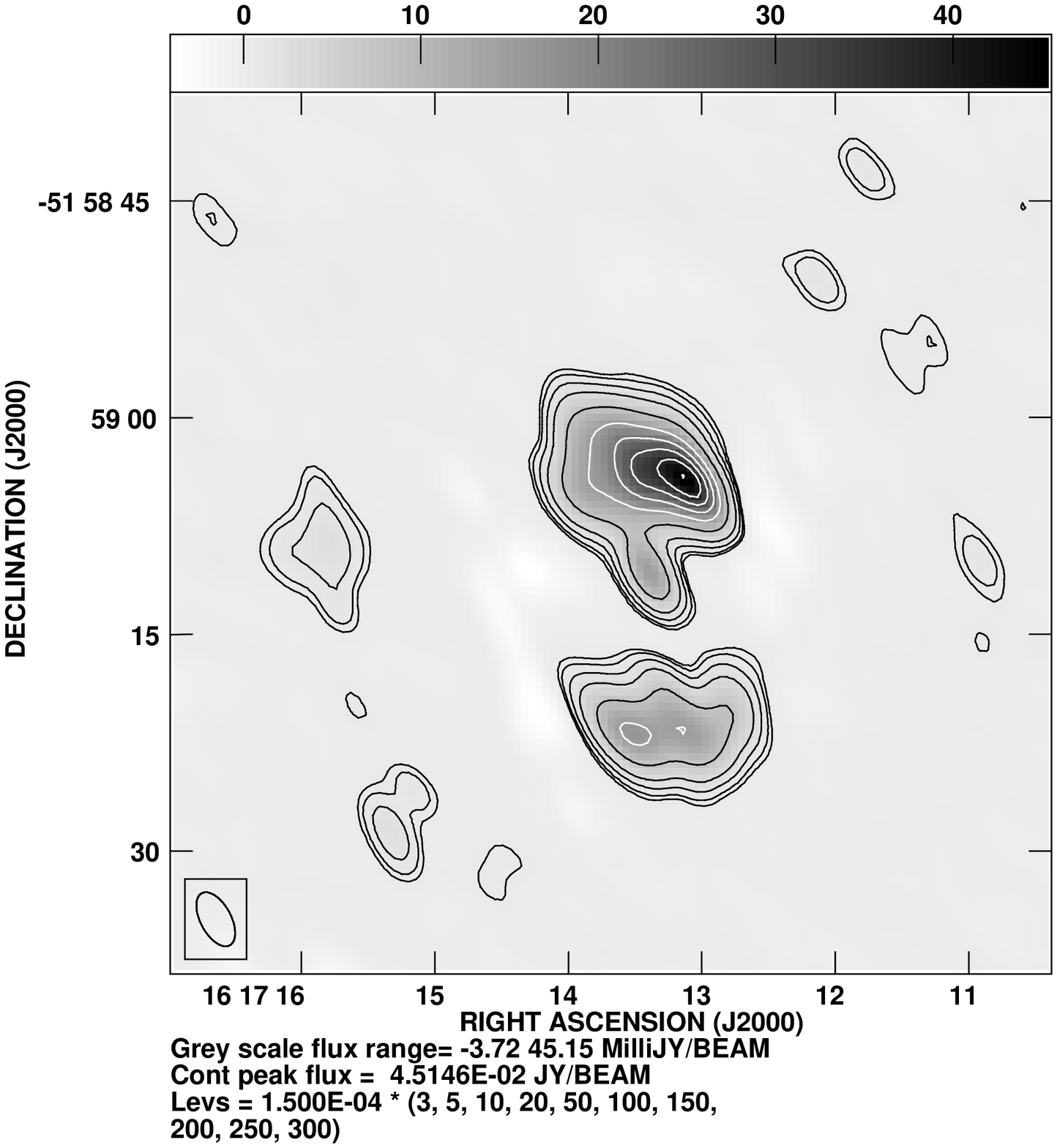}{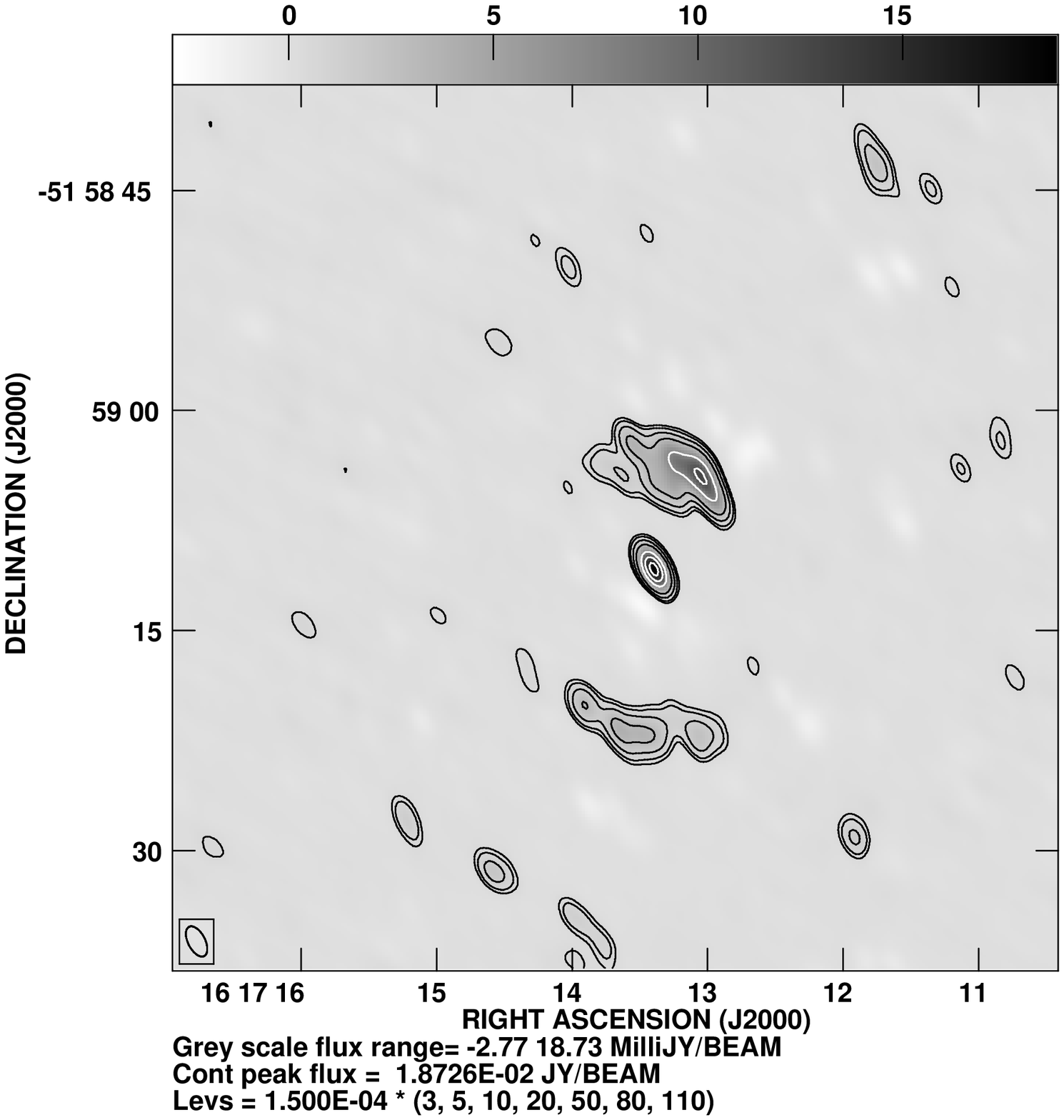}
\caption{Optical and radio images of Mz~3.  a) {\it HST} H$\alpha$
        image in logarithmic scales.  b) ATCA 6~cm (left) and 3.6~cm
        (right) radio continuum image in greyscale with contours
        superposed.  The contours start at $3\sigma$ with the noise
        level $\sigma = 1.5 \times 10^{-4}$ Jy/beam at 6~cm and
        $\sigma = 1.5 \times 10^{-4}$ Jy/beam at 3.6~cm.  The radio
        core and bipolar lobes coincide with the optical stellar core
        and the brightest structures of the optical lobes.}
\label{fg:mz3}
\end{center}
\end{figure}

  Of the five objects observed by ATCA, He~2-36 shows only extended
  emission, He~2-25 and He~2-84 shows only a compact central source,
  Mz~3 has both, and no emission was detected from Th~2-B.
  Figures~\ref{fg:he2-25} to \ref{fg:mz3} present the radio and
  optical images of the four objects with detections.  For the three
  objects with radio core detections, their optical images all show a
  compact stellar-like cores.  The optical images of He~2-25 were
  obtained from \cite{lee07}, those of Mz~3 from the {\it HST}
  archive, and those of He~2-36 and He~2-84 from SSO 2.3m telescope
  observations.  As before, images are oriented with north up, east to
  the left.

\subsubsection{He 2-25}

  The optical images of He~2-25 display a bright central source and a
  pair of very faint bipolar lobes extending toward north and south
  (Figure \ref{fg:he2-25}).  The radio emission of this object is only
  detected at 3.6~cm, which shows a central unresolved source
  coinciding with the optical core.

\subsubsection{He 2-36}

  The optical images of He~2-36 display a bright central source and an
  equatorial torus with limb brightening arms extending clockwise from
  the torus edges toward the north and south.  The brightening
  resembles an S-shape point symmetrical distribution.  We presented
  the H$\alpha$ image in two different scales in order to show the
  detailed structures (Figure \ref{fg:he2-36}).  The nebula possesses
  a more moderate degree of bipolarity (lobes are less collimated)
  than other butterfly nebulae observed in this work.  The radio
  images only show emission from the equatorial band.  Although the
  radio images show some interesting features, these are not relevant
  to current investigation.

\subsubsection{He 2-84}

  The H$\alpha$ and [N~{\footnotesize II}] images of He~2-84 show an
  optical core with an equatorial torus.  The two lobes extending at
  right angles to this torus toward the northeast and southwest appear
  to have a point symmetrical geometry.  The H$\alpha$ image is
  displayed in two different scales in order to show the detailed
  structures (Figure \ref{fg:he2-84}).  The lobes are very faint and
  cannot be seen in the [O~{\footnotesize III}] image.  The radio
  continuum images only show a central unresolved bright source, which
  coincides with the optical core.

\subsubsection{Mz 3}

  The optical images of Mz~3 show a bright core, two approximately
  spherical bipolar lobes, and even more extensive filamentary
  nebulosities on either side of the central star can be seen from its
  H$\alpha$ and [N~{\footnotesize II}] images.  Only the H$\alpha$
  images is presented in Figure \ref{fg:mz3}. The total size of the
  nebula spans over 50$''$ on the sky.  The radio images of Mz~3 show
  emission from the central source and the two lobes, similar to the
  radio images of M~2-9 \citep{kwok85}.  The radio central source and
  the two lobes coincide with the optical core and the brightest
  structures of the optical lobes.

\subsubsection{Spectral Indices of the Radio Cores}

  He~2-25 has a compact radio core detected only at 3.6~cm.  The fact
  that no detection at 6~cm has already indicated that this core is
  optically thick.  Using the detection limit at 6~cm, a lower limit
  of spectral index is estimated to be 1.25.  However, a more precise
  evaluation of the spectral index of He~2-25 has to wait for
  observations at shorter wavelengths.

  He~2-84 and Mz~3 have radio cores detected both at 3.6~cm and 6~cm,
  suitable for spectral index determination.  For He~2-84, we applied
  the task IMFIT to the clean map and UVFIT to the visibility data to
  determine the flux densities of its radio core assuming the radio
  core to have a two-dimensional Gaussian structure.  For Mz~3, the
  flux density was obtained only for the bright compact core in the
  CLEAN map.  This core can be seen in Figure \ref{fg:mz3}.  The total
  flux density was obtained by summing up the flux densities of the
  CLEAN components in the radio core region, since the convolution of
  the Clean components with the CLEAN beam might result in the
  confusion of the core emission and nebular emission.  The results
  are listed in Table~\ref{tb:atca-flux} with 1$\sigma$ uncertainties.
  Even with only clean components in the core region considered, it
  seems only part of the diffuse emission is excluded.  An analysis of
  this is given later.

\begin{figure}
\plottwo{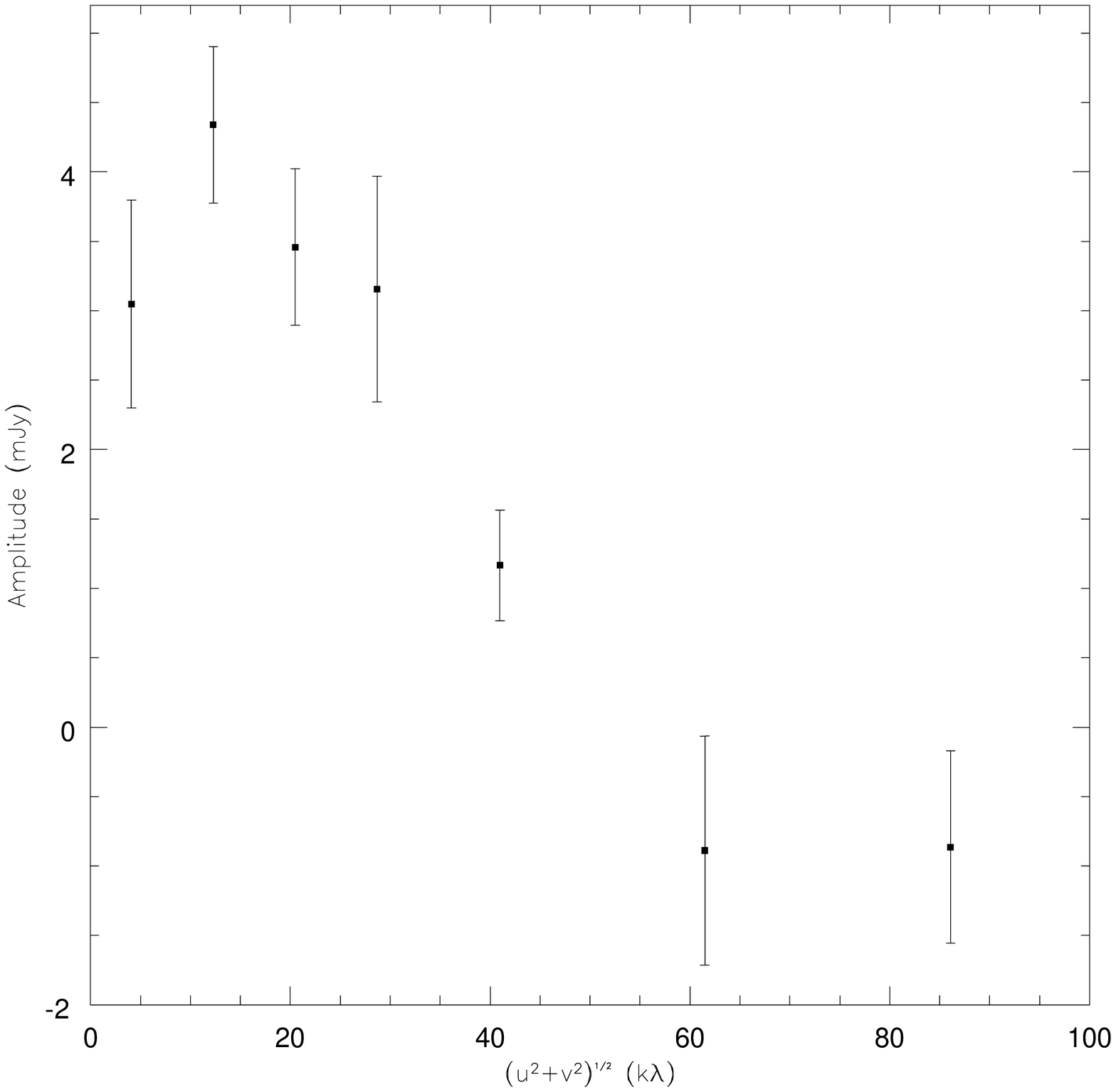}{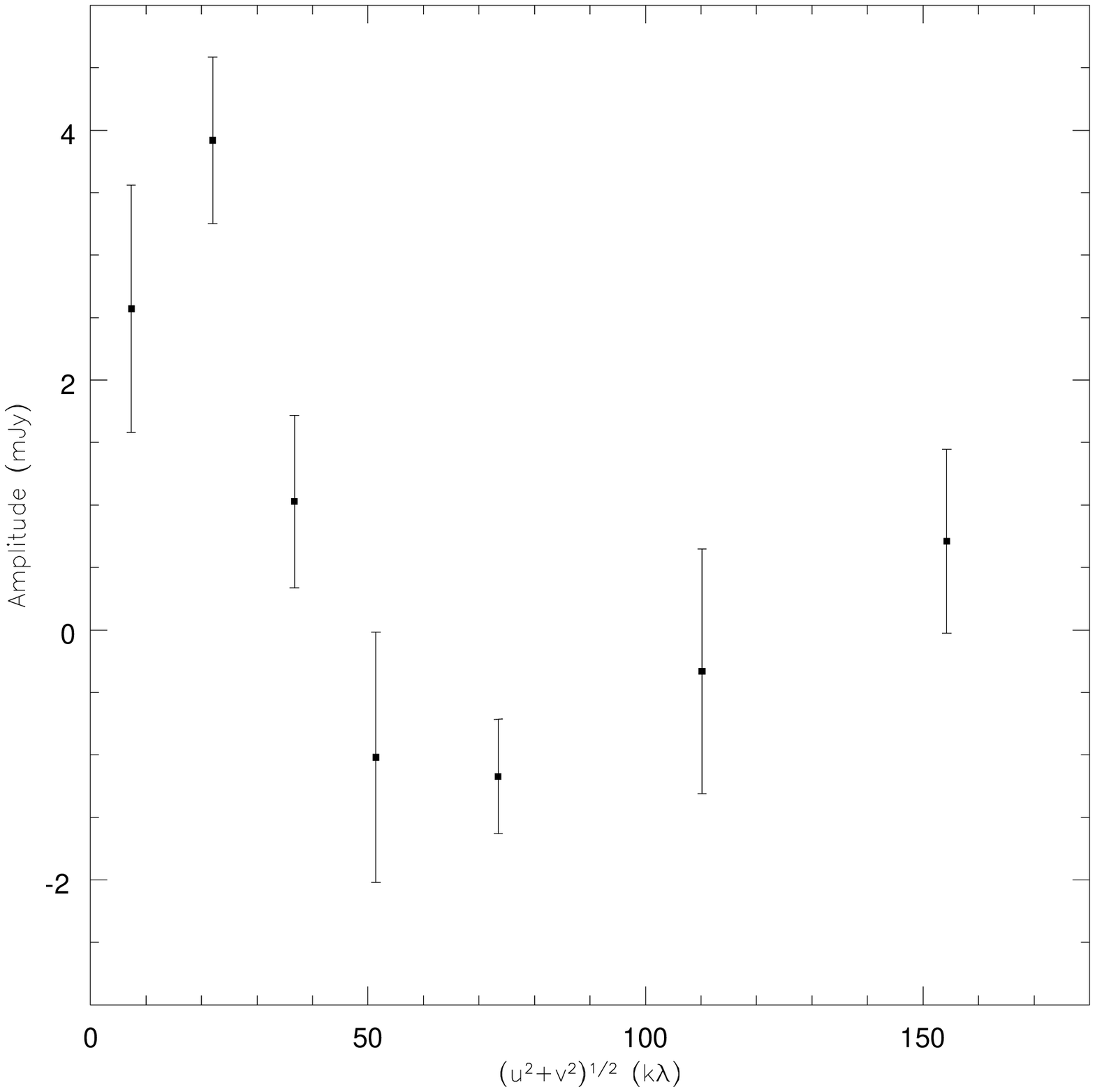}
\caption{Amplitude vs. {\it uv}-distance for He~2-84 at 6 cm (left)
         and 3.6 cm (right).  The real component of the ampplitudes
         are averaged in several {\it uv} bins.  The binning size is
         wider in the longer {\it uv}-distance in order to increase
         the signal to noise ratio at longer baselines.}
\label{fg:he2-84.uv-amp}
\end{figure}

  The radio core of He~2-84 has a spectral index significantly more
  negative than the value of $-0.1$ expected for optically thin
  free-free emission.  This suggests that the emission of the observed
  radio core is dominated by more extended emission that is, as
  expected, better resolved out at shorter than at longer wavelengths.
  Therefore the extended emission makes a stronger contribution at
  6~cm than at 3.6~cm, resulting in a spectral index that is even more
  negative than expected for optically thin free-free emission.  To
  illustrate the effects of the extended emission on the flux
  measurements, Figure \ref{fg:he2-84.uv-amp} plots amplitude vs. {\it
  uv}-distance for He 2-84 at 6 cm and 3.6 cm.  The real component of
  the amplitudes have been averaged in several {\it uv} bins.  The
  binning size is wider at longer {\it uv}-distance in order to
  increase the signal to noise ratio at longer baseline.  As shown in
  the figure, the amplitude decreases with increasing {\it
  uv}-distance which clear demonstrates that an extended component is
  present.

  In order to mitigate against contamination from the diffuse nebular
  emission, we excluded the short baselines ($<$ 50 k$\lambda$) that
  pick up the extended nebular component.  We also excluded the
  longest baselines at 3.6~cm in order to provide the same angular
  resolution as at 6~cm.  Because of the noise increases caused by
  removing {\it uv} data, the visibility and map fits did not yield
  reliable results.  Consequently, flux densities were derived by
  summing the flux densities of the CLEAN components.  The results are
  listed in Table~\ref{tb:atca-uv}, which indeed shows that the
  extended emission constitutes $\sim$ 90\% of the flux at 6~cm, and
  $\sim$ 80\% of the flux at 3.6~cm for He~2-84.  A similar analysis
  was also performed on Mz~3, since some contamination from the
  extended emission is also possible.  The results listed in
  Table~\ref{tb:atca-uv} show that for Mz~3, the flux contributed from
  the extended emission is relatively small at 6~cm, and there is
  virtually no contamination at 3.6~cm.

  The differences between the core flux densities and resultant
  spectral indices obtained using the different {\it uv} distances
  suggest that for both He~2-84 and Mz~3, the emission comes from two
  sources, one compact and one diffuse.  In both cases, the more
  careful analysis incorporating the same {\it uv}-coverage (Table
  \ref{tb:atca-uv}) indicates that an optically thick radio core has
  been detected.  A separate ATCA investigation of the radio
  properties of Mz~3 by \cite{bains04} also concluded that Mz~3 has an
  optically thick core.

\begin{figure}
\plottwo{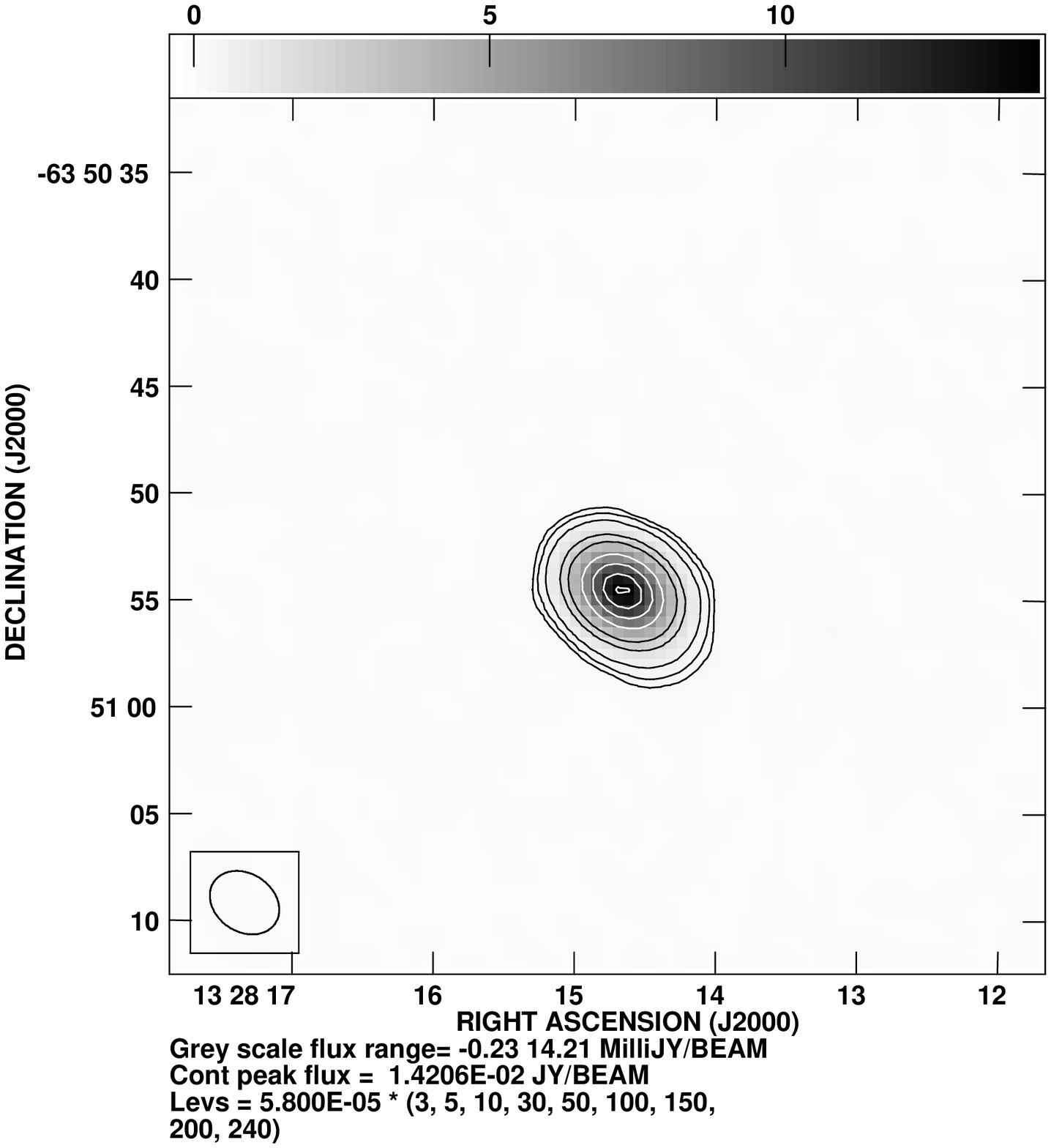}{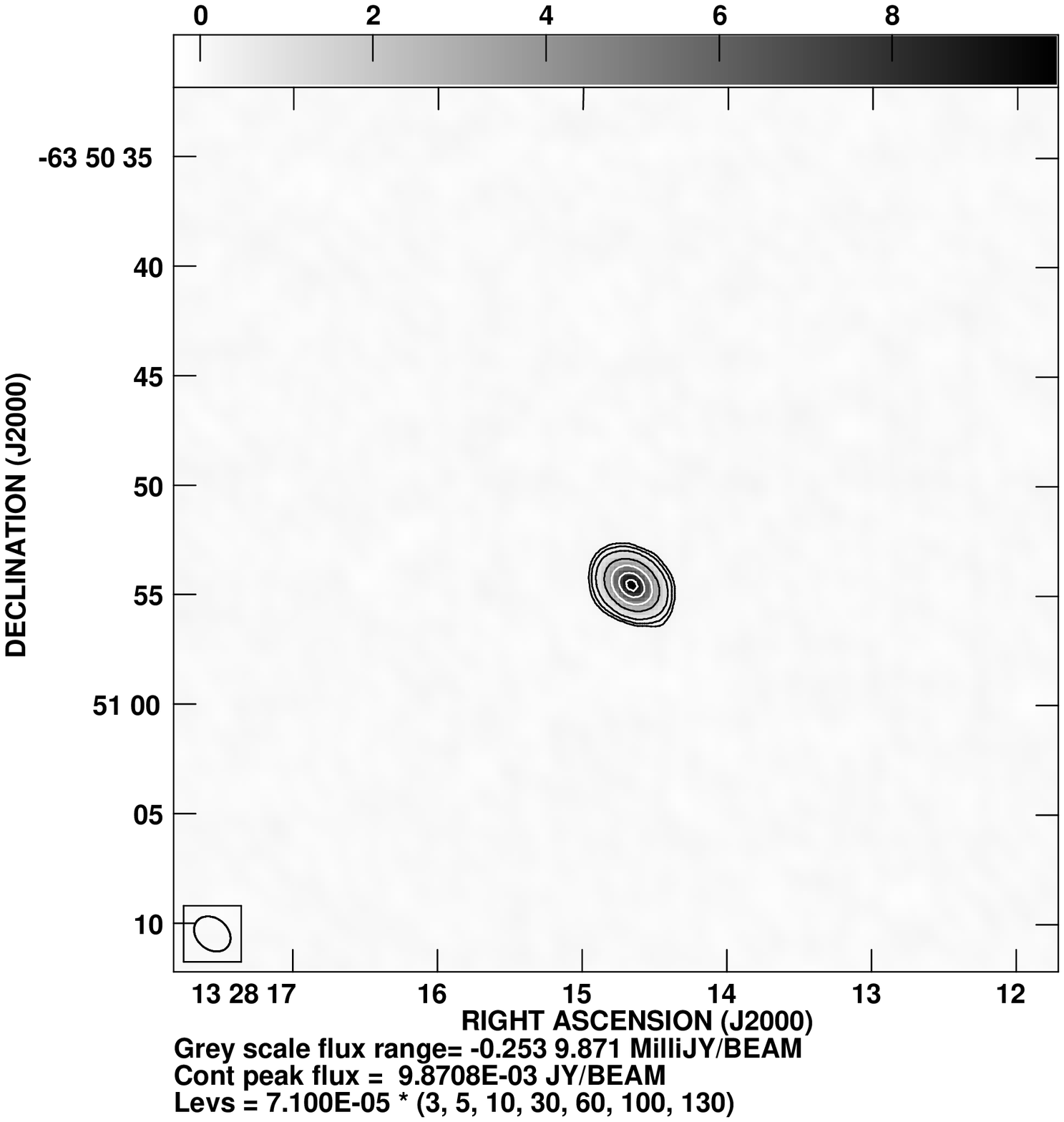}
\caption{Images of a background radio source near Th~2-B at 6~cm (left)
        and 3.6~cm (right).}
\label{fg:brs}
\end{figure}

  It is worthwhile mentioning that we did find a compact radio source
  both in 6~cm and 3.6~cm inside the primary beam of the Th~2-B
  observation as shown in Figure~\ref{fg:brs}.  However, after
  carefully examining the position of this source with the known
  objects in the field using
  SIMBAD\footnote{http://simbad.u-strasbg.fr/}, we determined that
  this is a background radio source not related to Th~2-B, since it is
  about 3$'$ away from the optical postion of Th~2-B.  This background
  radio source has a spectral index of $-1$, characteristic of
  nonthermal radio emission.  It is probably a quasar.

\section{Discussion}
\label{sec:discussion}

\subsection{The Mass Loss Rate}
\label{sec:mdot}

  The VLA and ATCA radio flux densities obtained in the previous
  sections can be used to estimate the mass loss rates of the nebulae,
  following the procedure in \cite{wright75}.  \citeauthor{wright75}
  gave the mass loss rate for a spherical wind with a spectral index
  of 0.6 as
\begin{equation}
  \dot{M} = 0.095 \frac{\mu_e v S_{\nu}^{3/4} D^{3/2}}{Z x_e^{1/2}
  g^{1/2} \nu^{1/2}}~~~~ {\rm M_{\odot}~ yr^{-1}}~,
\label{eq:mdot}
\end{equation}
  where {\it v} is the velocity of the mass loss flow in km~s$^{-1}$,
  $S_{\nu}$ is the flux observed at frequency $\nu$ in Jy, $\nu$ is in
  Hz, {\it D} is the distance in kpc, {\it Z} is the nuclear charge,
  $\mu_e$ is the mean atomic weight per electron, $x_e$ is the ratio
  of electron density to the number density of ionized H, i.e., $n_e =
  x_e n_p$, and {\it g} is the Gaunt factor, which can be approximated
  in the radio regime as
\begin{equation}
  g (\nu, T) = \frac{\sqrt{3}}{\pi} \left\{ 17.7 +  \ln \left[
  \frac{ (T_e/{\rm K})^{3/2}}{(\nu/{\rm Hz}) Z} \right] \right\}~,
\end{equation}
  where $T_e$ is the electron temperature \citep[p. 95]{osterbrock89}.

  The constants $x_e$ and $\mu_e$ can be found by assuming the He to H
  number ratio {\it y} and the fraction of He that is singly ionized
  $y'$, since the contributions from heavy elements are negligible due
  to their low abundance.  The singly ionized He to H ratio is $y'y$
  and the doubly ionized He to H ratio is $(1-y')y$ if there is no
  neutral He.  The ratio of electron density to the number density of
  ionized H becomes $x_e=1+yy'+2y(1-y')$, and the mean atomic weight
  per electron is
\begin{equation}
  \mu_e = \frac{1+4y}{1+yy'+2y(1-y')}
\end{equation}
  For $y=0.11$ and $y'=0.5$, $x_e$ and $\mu_e$ have values of 1.165
  and 1.236, respectively.

  Assuming $T_e=4000$ K \citep[the temperature of free electrons are
  mainly contributed by ionization of hydrogen,][]{reid97} and $Z=1$,
  we calculate the mass loss rate as a function of distance and wind
  velocity using the flux densities found in our observations.  Due to
  the large uncertainties of the mass loss rate parameters, the
  differences from the flux densities at different wavelengths for the
  same object are nebligible.  Therefore we give a best estimate of
  the mass loss rate for each object in Table~\ref{tb:mass}.

  To calculate the actual mass loss rates, the wind velocities and the
  distances to the nebulae are needed.  Assuming a typical wind
  velocity of 1000 km s$^{-1}$, and using the distances available from
  the literature as given in Table \ref{tb:mass}, we obtain $\dot{M}
  \sim 7 \times 10^{-6}~{\rm M_{\odot}~ yr}^{-1}$ for He~2-25,
  $\dot{M} \sim 10^{-5}~{\rm M_{\odot}~ yr}^{-1}$ for He~2-84,
  $\dot{M} \sim 3 \times 10^{-5}~{\rm M_{\odot}~ yr}^{-1}$ for 19W32,
  $\dot{M} \sim 9 \times 10^{-5}~{\rm M_{\odot}~ yr}^{-1}$ for M~1-91
  and $\dot{M} \sim 2 \times 10^{-4}~{\rm M_{\odot}~ yr}^{-1}$ for
  Mz~3, respectively.  The mass loss rate for Mz~3 is consistent with
  that derived by \cite{bains04}.  They found $\dot{M} \sim 5-30
  \times 10^{-5}~{\rm M_{\odot}~ yr}^{-1}$ using the range of $D$
  available in the literature.

  We also used the Eq.\ (11) in \cite{wright75} to calculate the size
  of the emitting source, under the spherical wind assumption.  The
  size equation, when incorporated Eq.\ (\ref{eq:mdot}), reduced to
\begin{eqnarray}
  r(\nu) &=& 5.83 \times 10^{27} S_{\nu}^{1/2} T^{-1/2} \nu^{-1} D
  ~~~{\rm cm} \nonumber \\
         &=& 3.9 \times 10^{11} S_{\nu}^{1/2} T^{-1/2} \nu^{-1}
  ~~~{\rm arcsec}~,
\label{eq:size}
\end{eqnarray}
  where $r$ corresponds to the radius of the emitting region where the
  optical depth $\tau_{\nu}=0.244$.  For the unresolved sources,
  Eq.~(\ref{eq:size}) predicts a source size smaller than the
  synthesized beam.  For the two resolved sources 19W32 and M~1-91, we
  listed the measured dimensions along the resolved axis and the
  predicted source sizes in Table \ref{tb:vla-size}.  The predicted
  sizes are about 4 or 5 times smaller than the measured dimensions.
  It is possible that we underestimated the predicted size by giving
  the wrong assumption of temperature in Eq.~(\ref{eq:size}).
  However, to make the predicted size consistent with the measured
  dimension, the temperature has to be unrealistically small ($\sim$
  200 K), in which case the winds are not ionized.  Therefore, we
  concluded that the predicted sizes are smaller is because these
  sources are elongated, possibly by collimated winds (jets).

  If we approximate a collimated source as an ellipse with major and
  minor axes of $a$ and $b$, and the collimated wind has an emission
  area as the same as a spherical wind, then
\begin{equation}
  a = r \sqrt{2/\theta_0} \ .
\end{equation}
  where $\theta_0=2b/a$ is the opening angle of the jet in radians.
  Using the measured dimension of the resolved axis as $a$, we can
  derive the opening angle of the collimated jet for 19W32 and M~1-91
  as listed in Table \ref{tb:vla-size}.  The opening angle can then be
  used to revise the mass loss rate.  As pointed out by
  \cite{reynolds86}, a collimated wind with a much lower mass loss
  rate can easily produce the same radio intensity as an uncollimated
  wind with a much higher mass loss rate, with the relationship for a
  spectral index of 0.6 is
\begin{equation}
  \frac{\dot{M}{\rm (jet)}}{\dot{M}{\rm (spherical~wind)}} = 0.20
  \theta_0 (\sin i)^{-1/4}
\end{equation}
  where {\it i} is the inclination of the major axis of the wind to
  our line of sight.  Assuming $i=90^{\circ}$, which gives a lower
  limit of the ratio, we derived the ratio is $\sim0.02$ in both
  cases.  Thus we revised the mass loss rates of 19W32 and M~1-91 to
  $\dot{M} \sim 10^{-6}~{\rm M_{\odot}~yr}^{-1}$ for collimated winds,
  which are about 50 times lower than assuming spherical symmetry.
  From the analysis of central-star wind P-Cygni profiles observed
  with the {\it International Ultraviolet Explorer}, the mass loss
  rates from central stars of PNs are $10^{-9}-10^{-7}$ M$_\odot$
  yr$^{-1}$ \citep{perinotto83}.  These values are consistent with
  theoretical values expected from winds driven by radiation pressure
  on ionized gas \citep{pau03}.  The mass loss rates we inferred are
  at least an order of magnitude higher than expected.  Such high mass
  loss rates can be explained if the central stars of these nebulae
  are symbiotic stars, as in the case suspected for M~2-9 whose
  inferred mass loss rate is $\sim 6 \times 10^{-6}~{\rm
  M_{\odot}~yr}^{-1}$ \citep{lim03}.  Similarly, \cite{bains04} also
  favored a binary companion scenario instead of the superwind of a
  single progenitor to explain the observational properties of Mz~3.

\subsection{PNs or Symbiotic Stars?}

  Symbiotic stars are interacting binary systems, typically consisting
  of a white dwarf that accretes material from the wind of a cool red
  giant companion, sometimes forming a disk around the white dwarf.
  The radiation from the hot star partially ionizes the wind from the
  cool star.  A spectrum of the two together thus shows a combination
  of absorption lines from the surface of the cool red giant and
  emission lines from the nebula.  First noticed by \cite{merrill19},
  \cite{plaskett28}, and \cite{merrill32} as stars with combination of
  spectra, these objects were named symbiotic stars by
  \cite{merrill58}.

  The possibility of some well-known PNs can in fact be symbiotic
  stars was discussed by \cite{kwok03b} and \cite{corradi03}.  With
  the discovery of increasing number of PNs with binary nuclei, the
  simplistic notion that PNs have single central stars and symbiotics
  are binaries is no longer viable.  It is often difficult to
  distinguish the two classes of objects by morphology, spectrum, or
  spectral energy distribution.  Only in cases where periodic
  photometric variability as a result of pulsation, or the
  observations of photospheric molecular absorption band in the
  spectrum, that one can be certain that we are dealing with a
  symbiotic system.  However, it is difficult to rule out that certain
  PNs, including those under study here, are NOT symbiotic stars.
  Even the evolutionary status of extremely well-studied objects such
  as M~2-9 is still not certain.

  If these nebulae with radio cores are indeed symbiotic star systems,
  the ionized wind we measured would originate from the white dwarf
  companion.  The mass loss rate of this wind does not directly tell
  the mass loss rate of the AGB star, but they must be related since
  the accretion disk is material captured from the AGB star wind.  If
  there is a balance between accretion onto the disk and ejection to
  form a jet, then the mass loss rate of the AGB star must be at least
  as high as the mass loss rate of the jet.  Assuming a spherically
  symmetric outflow and only a fraction of the slow neutral wind from
  the AGB star is captured by the WD dwarf, the corresponding mass
  loss rate of the AGB star would be at least in excess of $10^{-4}$
  M$_\odot$ yr$^{-1}$.  Such values are seldom observed in AGB stars.
  This problem can be overcome if the AGB star wind is strongly
  concentrated in the binary orbital plane, as is predicted for
  symbiotic star systems with relatively small orbital separations.

  The remaining objects have no detected radio core emission because
  they are either too distant or too faint.  None of those with known
  distances (0.5 - 7 kpc) are farther than the most distant PN with a
  detected radio core (M~1-91, 7 kpc).  If all these objects have the
  same radio luminosity, we should be able to detect all of them.
  Therefore the more likely explanation is that these objects have
  intrinsically faint radio emission.  It is possible that those
  objects evolved from single stars thus do not produce strong radio
  emission.  Since the mass loss rate from the central stars of PNs
  are at least an order of magnitude smaller than that of the
  symbiotic systems, even if these nebulae do have ionized jets from
  the central stars, their radio emission would be too faint to be
  detected.  Thus radio observations may provide a way to separate
  symbiotic nebulae from genuine planetary nebulae.  This would also
  mean that there is more than one mechanism to give a nebula a
  narrow-waist shape.  For single stars, one of the more possible
  mechanisms for shaping nebulae is magnetic fields
  \citep[e.g.,][]{garcia99}.

\section{Summary}
\label{sec:sum-jet}

  We have used the VLA and ATCA to search for signatures of ionized
  winds -- optically thick radio cores -- in sixteen narrow-waist
  bipolar nebulae.  Eleven nebulae were observed with the VLA at
  1.3~cm and 0.7~cm with resolutions of 0.08$''$ and 0.04$''$,
  respectively.  This is the first time that the centers of these
  objects have been systematically studied at such high angular
  resolution at radio wavelengths.  Two objects, 19W32 and M~1-91,
  were found to exhibit a compact central source with a positive
  spectral index.  Three more objects examined with the VLA, NGC~6302,
  Hubble~5, and NGC 6537, exhibit extended radio emission that cannot
  be mapped properly at such short wavelengths.  No emission was
  detected for six other objects in the VLA survey.

  Five nebulae were observed with the ATCA at 6~cm and 3.6~cm with
  resolutions of 3$''$ and 1.5$''$, respectively.  He~2-25 was found
  to have a compact central source only at 3.6~cm.  He~2-84 and Mz~3
  were found to contain a compact central source at both wavelengths.
  When short baselines that are sensitive to extended nebular emission
  were excluded, we found that the spectral index of these two radio
  cores changes with {\it uv}-coverage, suggesting that an optically
  thick core has been detected.  In addition to its compact radio
  core, Mz~3 also exhibits extended nebulae emission that coincides
  with the two lobes seen in the {\it HST} optical images.  He~2-36
  also shows extended nebular emission but no compact core.  No radio
  emission was detected for Th~2-B in the ATCA study.

  For 19W32 and M~1-91, we found that the predicted source sizes are
  smaller than the measured dimensions, which we attribute to the
  radio cores being elongated, implying a collimated wind or jet.  The
  possibility that these radio cores represent COFs responsible for
  shaping the bipolar morphology of the nebulae is particularly
  intriguing.  The derived mass loss rates of the wind from the radio
  observations are high in comparison to typical fast winds from
  central stars of planetary nebulae, and one possible solution is
  that the winds are not spherical symmetric but instead are
  collimated.  If these winds are collimated by a binary system, then
  the radio imaging offers the best way to test such models.  The five
  objects with radio core detections all exhibit a compact
  stellar-like optical core.  This needs further investigation in
  order to understand the connection between the optical morphology
  and the radio ionized jets.

  We are planning follow-up observations of these radio cores with
  higher angular resolution in order to study the structures of the
  cores.  We have recently obtained observations of M~1-91 with the
  VLA A+PT (Pie Town) configuration.  The addition of the Pie Town
  antenna will increase the resolution of the VLA by a factor of two
  while keeping the full sensitivity of the VLA.  The new data will
  add to our knowledge of the core properties of M~1-91 once the
  analysis is completed.

\acknowledgements

  We thank the anonymous referee for constructive comments on the
  manuscript.  We are grateful to the local staff of the NRAO,
  Narrabri and Siding Spring Observatories for their support during
  observations at the VLA, ATCA and SSO.  The National Radio Astronomy
  Observatory is a facility of the National Science Foundation
  operated under cooperative agreement by Associated Universities,
  Inc.  The Compact Array is part of the Australia Telescope, which is
  funded by the Commonwealth of Australia for operation as a National
  Facility managed by the Commonwealth Scientific and Industrial
  Research Organisation.  The Siding Spring Observatory is operated by
  the Australian National University, Research School of Astronomy and
  Astrophysics.

  THL expresses gratitude to the University of Calgary and the
  Province of Alberta for various scholarships and a travel grant
  during her PhD.  This work was supported in part by a grant to SK
  from the Natural Sciences and Engineering Research Council of
  Canada.

\clearpage

\begin{deluxetable}{ccccccc}
\tablecaption{The properties of the
  selected PNs with the first 11 objects observed by the VLA and the
  last 5 objects observed by the ATCA.\label{tb:objects-jet}}
\tablehead{
\colhead{Object} & \colhead{PN G} & \colhead{RA} & \colhead{Dec} &
  \colhead{VNW} & \colhead{Optical core} & \colhead{Point/Mirror}
}
 \startdata
  IRAS 07131-0147 & \nodata & 07:15:42.6 & $-$01:52:42 & yes & yes & ? \\
  M 1-16 & $226.7+05.6$ & 07:37:18.6 & $-$09:38:48 & ? & ? & yes \\
  NGC 2818 & $261.9+08.5$ & 09:16:01.5 & $-$36:37:37 & ? & no & ? \\
  NGC 6302 & $349.5+01.0$ & 17:13:44.2 & $-$37:06:14 & yes & no & no \\
  19W32 & $359.2+01.2$ & 17:39:02.8 & $-$28:56:35 & yes & yes & ? \\
  HB 5 & $359.3-00.9$ & 17:47:56.8 & $-$29:59:53 & yes & yes & yes \\
  NGC 6537 & $010.1+00.7$ & 18:05:13.2 & $-$19:50:34 & yes & ? & yes \\
  M 3-28 & $021.8-00.4$ & 18:32:41.3 & $-$10:06:05 & yes & ? & yes \\
  M 1-91 & $061.3+03.6$ & 19:32:57.3 & $+$26:52:43 & yes & yes & yes \\
  M 2-48 & $062.4-00.2$ & 19:50:27.9 & $+$25:54:28 & yes & ? & no \\
  NGC 7026 & $089.0+00.3$ & 21:06:18.5 & $+$47:51:08 & ? & ? & yes \\ \tableline
  He 2-25 & $275.2-03.7$ & 09:18:01.3 & $-$54:39:29 & yes & yes & ? \\
  He 2-36 & $279.6-03.1$ & 09:43:25.6 & $-$57:16:56 & ? & yes & yes \\
  He 2-84 & $300.4-00.9$ & 12:28:46.7 & $-$63:44:35 & yes & yes & yes \\
  Th 2-B & \nodata & 13:28:38.2 & $-$63:49:42 & ? & yes & ? \\
  Mz 3 & $331.7-01.0$ & 16:17:12.6 & $-$51:59:08 & yes & yes & yes  \\
\enddata
\end{deluxetable}

\begin{deluxetable}{cccc}
\tablecaption{Optical observing parameters\label{tb:filter}}
\tablehead{ \colhead{Filter Name} & \colhead{Center $\lambda$} &
\colhead{Bandwidth $\Delta \lambda$} & \colhead{FWHM} \\
 & \colhead{(\AA)}& \colhead{(\AA) }& \colhead{($''$)} 
}
\startdata 
 H$\alpha$ & 6568 & 8 & 2.2 \\ 
 $[$\ion{N}{2}$]$ & 6588 & 8.8 & 2.4 \\ 
 $[$\ion{O}{3}$]$ & 5007 & 10 & 2.4 \\
\enddata
\end{deluxetable}

\begin{deluxetable}{ccccccc}
\tablecaption{Results of the VLA and ATCA Surveys.\label{tb:result}}
\tablehead{
\colhead{Object} & \multicolumn{2}{c}{Detection limit} &
\multicolumn{2}{c}{Synthesized beam size} & \colhead{Type of
  Detection} \\
& \multicolumn{2}{c}{(Jy)} & \multicolumn{2}{c}{($''$)} & 
}
 \startdata
                  & 1.3 cm & 0.7 cm & 1.3 cm       & 0.7 cm \\
  IRAS 07131-0147 & 3.0E-4 & 7.2E-4 & 0.11 x 0.079 & 0.080 x 0.041 & \nodata \\
  M 1-16          & 3.3E-4 & 7.8E-4 & 0.13 x 0.077 & 0.080 x 0.042 & \nodata \\
  NGC 2818        & 5.1E-4 & 1.5E-3 & 0.26 x 0.082 & 0.16 x 0.042 & \nodata \\
  NGC 6302        & 6.6E-4 & 4.2E-3 & 0.26 x 0.083 & 0.17 x 0.041 & extended \\
  19W32           & 3.6E-4 & 9.9E-4 & 0.19 x 0.082 & 0.12 x 0.040 & compact \\
  HB 5            & 3.9E-4 & 9.6E-4 & 0.20 x 0.078 & 0.12 x 0.042 & extended \\
  NGC 6537        & 3.9E-4 & 9.9E-4 & 0.16 x 0.077 & 0.094 x 0.043 & extended \\
  M 3-28          & 2.9E-4 & 7.2E-4 & 0.14 x 0.079 & 0.080 x 0.046 & \nodata \\
  M 1-91          & 3.0E-4 & 8.1E-4 & 0.10 x 0.083 & 0.061 x 0.045 & compact \\
  M 2-48          & 3.0E-4 & 7.5E-4 & 0.10 x 0.084 & 0.060 x 0.045 & \nodata \\
  NGC 7026        & 3.3E-4 & 8.1E-4 & 0.096 x 0.091 & 0.068 x 0.042 & \nodata \\ \tableline
                  & 6 cm   & 3.6 cm & 6 cm        & 3.6 cm \\
  He 2-25         & 3.3E-4 & 1.9E-4 & 3.74 x 2.71 & 1.98 x 1.42 & compact \\
  He 2-36         & 1.8E-4 & 1.8E-4 & 3.59 x 2.87 & 1.88 x 1.52 & extended \\
  He 2-84         & 1.7E-4 & 2.1E-4 & 3.45 x 3.02 & 1.76 x 1.60 & compact \\
  Th 2-B          & 1.7E-4 & 2.1E-4 & 3.47 x 2.71 & 1.82 x 1.42 & \nodata \\
  Mz 3            & 4.5E-4 & 4.5E-4 & 4.09 x 2.11 & 2.24 x 1.16 & compact + extended \\
\enddata
\end{deluxetable}

\begin{deluxetable}{ccccc}
\tablecaption{Flux densities for compact cores of 19W32 and
  M~1-91\label{tb:vla-flux}}
\tablehead{
\colhead{Object} & \colhead{Fitting Process} & \colhead{1.3 cm} &
  \colhead{0.7 cm} & \colhead{Spectral Index} \\ 
& & \colhead{(mJy)} & \colhead{(mJy)}
}
\startdata
    & Visibility fit & 3.58 $\pm$ 0.18 & 6.18 $\pm$ 0.69 & 0.81 $\pm$
    0.18 \\
    \raisebox{1.5ex}[0pt]{19W32} & Map fit & 3.48 $\pm$ 0.17 & 4.76
    $\pm$ 0.57 & 0.47 $\pm$ 0.19 \\ \tableline
    & Visibility fit & 2.97 $\pm$ 0.16 & 5.11 $\pm$ 0.39 & 0.79
    $\pm$ 0.14 \\
    \raisebox{1.5ex}[0pt]{M 1-91} & Map fit & 2.65 $\pm$ 0.15 & 3.97
    $\pm$ 0.38 & 0.64 $\pm$ 0.17 \\ 
\enddata
\end{deluxetable}

\begin{deluxetable}{ccccc}
\tablecaption{Flux densities for compact cores of He~2-84 and
  Mz~3\label{tb:atca-flux}}
\tablehead{
\colhead{Object} & \colhead{Fitting Process} & \colhead{6 cm} &
  \colhead{3.6 cm} & \colhead{Spectral Index} \\ 
& & \colhead{(mJy)} & \colhead{(mJy)}
}
\startdata
    He 2-25 & Map pixel sum & $\leq$ 0.33 & 0.69 $\pm$ 0.06 &
  $\gtrsim + 1.25$\\ \tableline 
    & Visibility fit & 12.1 $\pm$ 0.16 & 9.2 $\pm$ 0.16 & $-$0.47
    $\pm$ 0.04 \\
    \raisebox{1.5ex}[0pt]{He 2-84} & Map fit & 8.6 $\pm$ 0.06 & 7.7
    $\pm$ 0.08 & $-$0.18 $\pm$ 0.02 \\ \tableline
    Mz 3  & Map pixel sum & 15.3 $\pm$ 0.15 & 18.3 $\pm$ 0.15 & $+$0.35
    $\pm$ 0.02
\enddata
\end{deluxetable}

\begin{deluxetable}{ccccc}
\tablecaption{Compact core fluxes vs. uv-range for He~2-84 and
  Mz~3\label{tb:atca-uv}} 
\tablehead{
\colhead{Object} & \colhead{{\it uv}-range} & \colhead{6 cm} &
  \colhead{3.6 cm} & \colhead{Spectral Index} \\ 
& \colhead{(k$\lambda$)} & \colhead{(mJy)} & \colhead{(mJy)}
}
\startdata
    & 50-100 & 1.83 $\pm$ 0.14 & 2.60 $\pm$ 0.10 & $+$0.60 $\pm$ 0.15
    \\
    He 2-84 & 65-100 & 1.39 $\pm$ 0.16 & 1.89 $\pm$ 0.12 & $+$0.52
    $\pm$ 0.22 \\
    & 75-100 & 1.08 $\pm$ 0.18 & 1.42 $\pm$ 0.15 & $+$0.47 $\pm$ 0.34
    \\ \hline
    & 45-100 & 13.3 $\pm$ 0.11 & 19.5 $\pm$ 0.11 & $+$0.65 $\pm$ 0.02
    \\
    \raisebox{1.5ex}[0pt]{Mz 3} & 55-100 & 10.8 $\pm$ 0.14 & 20.3
    $\pm$ 0.11 & $+$1.07 $\pm$ 0.02 
\enddata
\end{deluxetable}

\begin{deluxetable}{cccc}
\tablecaption{Mass loss rates and distances of the five objects with
  radio core detection\label{tb:mass}}
\tablehead{
\colhead{Object} & \colhead{$\dot{M}_{vD}$\tablenotemark{\dagger}} & \colhead{Distance} &
  \colhead{D. Reference} \\
& \colhead{($M_{\odot} {\rm yr}^{-1})$} &  \colhead{(kpc)} 
}
\startdata
    19W32 & $\sim 5 \times 10^{-9}$ & 3.0 & 1 \\
    M 1-91 & $\sim 5 \times 10^{-9}$ & 7.0 & 2 \\
    He 2-25 & $\sim 2 \times 10^{-9}$ & 2.1 & 3 \\
    He 2-84 & $\sim 5 \times 10^{-9}$ & 1.6 & 2 \\
    Mz 3 & $\sim 3 \times 10^{-8}$ & 3.3 & 4 \\
\enddata
\tablenotetext{\dagger}{$\dot{M} (M_{\odot} {\rm yr}^{-1})=\dot{M}_{vD}
    \times (v{\rm /km~s}^{-1})(D{\rm /kpc})^{3/2}$}
\tablerefs{1. \cite{kohoutek82}; 2. \cite{maciel84}; 3. \cite{phillips04};
  4. \cite{vanderveen89}.}
\end{deluxetable}

\begin{deluxetable}{cccccc}
\tablecaption{The measured dimensions and predicted sizes of 19W32 and
  M~1-91\label{tb:vla-size}}
\tablehead{
\colhead{Object} & \colhead{$\lambda$} & \colhead{Resolved Axis} &
  \colhead{P.A.\tablenotemark{a}} & \colhead{$r(\nu)$} &
  \colhead{$\theta_0$} \\  
 & \colhead{(cm)} & \colhead{(mas)} & \colhead{($^{\circ}$)} &
  \colhead{(mas)} & \colhead{($^{\circ}$)}
}
\startdata
    & 1.3 & 63 $\pm$ 11 & 51 $\pm$ 19 & 17 & 7.8 \\
    \raisebox{1.5ex}[0pt]{19W32} & 0.7 & 51 $\pm$ 10 &
    56 $\pm$ 46 & 10 & 4.3 \\ \tableline 
    & 1.3 & 77 $\pm$ ~7 & 86 $\pm$ ~6 & 14 & 4.0 \\
    \raisebox{1.5ex}[0pt]{M 1-91} & 0.7 & 34 $\pm$ ~6 &
    58 $\pm$ 12 & ~9 & 8.0\\
\enddata
\tablenotetext{a}{Positon angle of the resolved axis.}
\end{deluxetable}

\clearpage

\end{document}